\documentclass[11pt,a4paper]{article}
\usepackage{jcappub}
\usepackage{amssymb,amsmath}
\usepackage{graphics}

\newcommand{\beq}{\begin{equation}}
\newcommand{\eeq}{\end{equation}}
\newcommand{\bea}{\begin{eqnarray}}
\newcommand{\eea}{\end{eqnarray}}
\newcommand{\vc}[1]{{\textbf{#1}}}
\newcommand{\mc}[1]{\mathcal{#1}}
\newcommand{\cH}{\mathcal{H}}
\newcommand{\cS}{\mathcal{S}}

\author[a]{Kari Enqvist}
\author[a]{Shaun Hotchkiss}
\author[b]{Gerasimos Rigopoulos}

\affiliation[a]{University of Helsinki and Helsinki Institute of Physics, P.O.Box 64,
FIN-00014, University of Helsinki,
Finland}

\affiliation[b]{
Institute for Theoretical Particle Physics and Cosmology, RWTH Aachen, D -
52056, Germany}

\title{A gradient expansion for cosmological backreaction}

\abstract{We address the issue of cosmological backreaction from non-linear
structure formation by constructing an approximation for the time evolved metric of a dust dominated universe based on a gradient expansion. Our metric begins as a perturbation of a flat Friedmann-Robertson-Walker state described by a nearly scale invariant, Gaussian, power-law distribution, and evolves in time until non-linear structures have formed. After describing and attempting to control for certain complications in the implementation of this approach, this metric then forms a working model of the universe. We numerically calculate the evolution of the average scale factor in this model and hence the backreaction. We argue that, despite its limitations, this model is more realistic than previous models that have confronted the issue of backreaction. We find that the \emph{instantaneous} effects of backreaction in this model could be as large as $\sim10\%$ of the background. This suggests that a proper understanding of the \emph{cumulative} effects of backreaction could be crucial for precision cosmology and any future exploration of the dark sector.}

\begin{document}

\maketitle

\section{Introduction}

The issue of cosmological backreaction \cite{Buchert:1999er} has fueled a lively debate in the literature. The question is: can inhomogeneities in our universe backreact and affect the average dynamics within our causal horizon such that the observed acceleration can be attributed to their influence? If the answer is positive then the mystery of the cosmological constant will have found a solution
requiring no new physics but which will nevertheless demonstrate the subtlety of
gravitational phenomena. Even if the answer is negative, backreaction will be
operative at some level due to the non-linear nature of gravity and its effects
may be visible in the new generation of cosmological observations. Either way,
calculating its magnitude is an important cosmological question.

The problem with assessing the magnitude of this backreaction lies with the
complexities of the non-linear structures forming under gravity in the late
universe. For example, using second order perturbation theory falls short of
providing an answer since the effect is expected to emerge in the non-linear
regime where perturbation theory breaks down. Attempts to model the non-linear
structures involving toy models of voids and overdensities may be useful for
developing intuition but are nevertheless simplistic. N-body simulations imply that the effect is small but they are Newtonian with zero global backreaction by
construction; the backreaction is a purely relativistic effect \cite{Buchert:1995fz}. A strong
argument for the smallness of the effect uses the fact that even with non-linear
overdensities, the local metric perturbations and peculiar velocities are still much
smaller than unity (away from black holes) and that a perturbative FRW framework
should therefore hold. However, counterarguments have been put forward involving
subtleties in the choice of background employed in these treatments. For
references finding a a small backreaction see e.g. \cite{Ishibashi:2005sj,
Paranjape:2008jc, Baumann:2010tm, Alonso:2010zv, Green:2010qy, Mattsson:2010ky} while
arguments for significant backreaction can be found in e.g. \cite{Rasanen:2006kp,
Rasanen:2008it, Clarkson:2009hr, estimate, Collins:2010ea, Bochner:2011dr}. For recent articles and reviews on the subject with more extensive references  see \cite{Buchert:2011sx, arXiv:1103.2335, arXiv:1102.1449,arXiv:1103.5974, arXiv:1102.1015, arXiv:1105.0909, arXiv:1106.1693, arXiv:1103.2016, arXiv:1102.0408, Kolb, arXiv:1105.1886, arXiv:1005.0788}.

In this paper we discuss cosmological backreaction in a novel manner, with a fully relativistic framework for a universe that begins as a perturbed Friedmann-Robertson-Walker, $\Omega=1$, CDM universe. To be precise, we employ a
gradient expansion to express the metric as a series of terms with an increasing
number of spatial gradients and coefficients which are functions of proper time.
The initial conditions are the standard adiabatic and Gaussian post-inflationary
primordial perturbations. We use the synchronous gauge: our coordinate lines
comove with CDM particles and our time hypersurfaces are labeled by their proper
time. Of course the gradient series has to be truncated and thus does not
capture the developing non-linearities entirely realistically. However, we argue
that even if truncated this series can still provide a well motivated \emph{model} for the true geometry which can be made increasingly accurate in principle. Furthermore, it extends into the non-linear regime, describing the collapse of initial over-densities and the rarefaction of initial under-densities which go on to form the voids dominating the cosmological volume.\footnote{The synchronous gauge and a gradient expansion to study backreaction was first used in
\cite{Kolb:2005da} to argue that a non-perturbative approach is really needed to settle the issue. In the present work we use a truncated gradient series but also consider it in the non-perturbative regime, beyond its apparent radius of convergence. As explained in the text we argue that this can be used as a model for the metric of the universe which captures the evolution of over- and under-dense regions.}

Here is the outline of the rest of the paper: In the following section we obtain the series solution for the metric in a gradient expansion using a Hamilton-Jacobi formulation. This approach, first developed in e.g. \cite{Croudace:1993yt} and \cite{Stewart:1994wq}, simplifies the calculation significantly compared to a more straightforward expansion of the standard Einstein equations. Our treatment follows a slightly different logical development. Then, in section 3, we apply these results to the backreaction problem by numerically evaluating the evolution of the average scale factor and the backreaction parameter $Q$. We find that, up to certain qualifications which we explain, backreaction leads to non-negligible deviations from the unperturbed background model. Our results indicate that the effect might be relevant, indeed crucial, for precision cosmology and the exploration of the dark sector. We summarize and discuss our findings in section 4 where future directions are also laid out.

\section{The inhomogeneous CDM Universe as a series in spatial gradients}
The study of the backreaction of cosmological inhomogeneities indicates that relevant effects, if any, will necessarily be in the relativistic and non-linear regime, in the realm beyond cosmological perturbation theory. In this regime deviations from a homogeneous FRW Universe can be studied using an expansion in spatial gradients. Furthermore, calculations beyond the lowest quasi-homogeneous order can be drastically simplified via the application of such a gradient expansion in the Hamilton-Jacobi formulation of gravitational dynamics. We present this formulation below and use it to obtain the first terms in the gradient series for the metric. We then compare with an exact spherically symmetric solution to gauge the accuracy of the approximation and develop some intuition. The section ends with a description of how to apply this gradient expansion for evaluating the backreaction of cosmological inhomogeneities by controlling a number of complications that arise in such an application.

\subsection{Hamilton-Jacobi for CDM}\label{HJ}
The Hamilton-Jacobi approach requires a Hamiltonian formulation which in turn requires an action. The action for gravity and a non-relativistic matter fluid (dust) can be written as
\beq
\cS=\int d^4x\sqrt{-g}\left[\frac{1}{2\kappa} \,{}^{(4)}\!R-\frac{1}{2}\rho\left(g^{\mu\nu}\partial_\mu\chi\partial_\nu\chi+1\right)\right]
\eeq
where $\chi$ is a potential for the 4-velocity of the fluid, $U^\mu=-g^{\mu\nu}\partial_\nu\chi$, and $\rho$, the energy density, acts as a Lagrange multiplier whose variation ensures that $U^\mu U_\mu=-1$. Variation wrt $\chi$ gives the continuity equation $\nabla_\mu(\rho\partial^\mu\chi)=0$, while variation of the dust part of the action wrt $g^{\mu\nu}$ gives the usual energy momentum tensor $T_{\mu\nu} = \rho U_\mu U_\nu$. Gravity is described by the standard Einstein-Hilbert term.

Let us now use the ADM decomposition for the metric and develop a Hamiltonian formalism. The metric is written as
\beq
  g_{00}= -N^2+h_{ij}N^iN^j \,,\qquad
  g_{0i}= \gamma_{ij}N^j \,,\qquad
  g_{ij}= \gamma_{ij}
  \,,
 \label{ADM metric}
\eeq
with the inverse
\beq
  g^{00}= -\frac{1}{N^2} \,,\qquad
  g^{0i}= \frac{N^i}{N^2} \,,\qquad
  g^{ij}= \gamma^{ij}-\frac{N^iN^j}{N^2}
  \,.
 \label{ADM metric:i}
 \eeq
By defining the canonical momenta
\bea
\pi^{ij}&\equiv&\frac{\delta \cS}{\delta\dot{\gamma}_{ij}}=\frac{\sqrt{\gamma}}{2\kappa}\frac{E^{ij}-\gamma^{ij}E}{N}\,,\\
\pi^\chi&\equiv&\frac{\delta \cS}{\delta\dot{\chi}}=\rho\sqrt{\gamma}\sqrt{1+\gamma^{ij}\partial_i\chi\partial_j\chi}\,,
\eea
with
\beq
 E_{ij} = \frac{1}{2} \dot \gamma_{ij} - \nabla_{(i}N_{j)}
\,,\qquad
 E =  h^{ij}E_{ij}
\,,
 \label{Eij-E}
\eeq
we can bring the action to the canonical form
\beq\label{canonical1}
\cS=\int d^4x\left(\pi^\chi\frac{\partial\chi}{\partial t}+\pi^{ij}\frac{\partial \gamma_{ij}}{\partial t}-N\mathcal{U}-N_i\mathcal{U}^i\right)
\eeq
where
\beq
\mathcal{U} = \frac{2\kappa}{\sqrt{\gamma}}
                     \left(\pi_{ij}\pi^{ij}-\frac{\pi^2}{2}
                     \right) - \frac{\sqrt{\gamma}}{2\kappa}R+\pi^\chi\sqrt{1+\gamma^{ij}\partial_i\chi\partial_j\chi}
\eeq
and
\beq
\mathcal{U}_i=-2\nabla_k\pi_i^k+\pi^\chi\partial_i\chi
\eeq

As is well known \cite{Arnowitt:1962hi}, the action (\ref{canonical1}) does not define a conventional Hamiltonian system since solutions to the equations of motion must set the ``Hamiltonian'' to zero, $\mathcal{U}=0$, a fact which reflects the time reparametrization invariance of General Relativity. One can however obtain a conventional Hamiltonian formulation by using as a time parameter one of the scalar fields of the system \cite{Arnowitt:1962hi}; in this case the obvious choice is to use the hypersurfaces of $\chi$ as time hypersurfaces, setting $\frac{\partial\chi}{\partial t}=1$, $N=1$ and $\partial_i\chi=0$. Thus, the metric takes the form
\beq
ds^2=-dt^2+\gamma_{ij}(t,\vc{x})dx^idx^j
\eeq
and the spatial coordinate lines comove with the matter. We then impose the energy constraint $\mathcal{U}=0$ and from this equation determine the (non-zero) $\pi^\chi$ which now plays the role of the Hamiltonian density: $-\pi^\chi\equiv \cH$. In particular
\beq
\mathcal{U}=0 \Rightarrow \cH =\frac{2\kappa}{\sqrt{\gamma}}
                     \pi_{ij}\pi^{kl}\left(\gamma_{ik}\gamma_{jl}-\frac{1}{2}\gamma_{ij}\gamma_{kl}
                     \right) - \frac{\sqrt{\gamma}}{2\kappa}R\,,
\eeq
and the action becomes
\beq\label{canonical2}
\cS=\int dt d^3x\left(\pi^{ij}\frac{\partial \gamma_{ij}}{\partial t}-\cH+ 2N_i\nabla_k\pi^{ki}\right)\,.
\eeq
In this form it defines a constrained Hamiltonian system where the canonical momentum $\pi^{ij}$ is constrained to be covariantly conserved
\beq\label{cov-cons}
\nabla_k\pi^{ki}=0\,.
\eeq

Let us now apply the Hamilton-Jacobi approach to the Hamiltonian system (\ref{canonical2}). Writing
\beq
\pi^{ij}=\frac{\delta \cS}{\delta\gamma_{ij}}
\eeq
we obtain the Hamilton-Jacobi equation
\beq\label{H-J}
\frac{\partial \cS}{\partial t}+\int d^3x \left\{\frac{2\kappa}{\sqrt{\gamma}}
                    \frac{\delta \cS}{\delta\gamma_{ij}} \frac{\delta \cS}{\delta\gamma_{kl}} \left(\gamma_{ik}\gamma_{jl}-\frac{1}{2}\gamma_{ij}\gamma_{kl}
                     \right) - \frac{\sqrt{\gamma}}{2\kappa}R\right\}=0\,,
\eeq
which is a single partial differential equation for $\cS$ as a functional of $\gamma_{ij}$ and a function of $t$. Once $\cS[t,\gamma_{ij}]$ is determined the metric can be obtained from
\beq\label{dotgamma}
\frac{\partial\gamma_{ij}}{\partial t}=\frac{2}{\sqrt{\gamma}}\frac{\delta\cS}{\delta\gamma_{kl}}\left(2\gamma_{ik}\gamma_{jl}-\gamma_{ij}\gamma_{kl}\right)
\eeq

Let us now turn to the remaining constraint (\ref{cov-cons}). It will be automatically satisfied if $\cS=\int d^3x \sqrt{\gamma}\,\mathcal{F}(t,\gamma_{ij})$ where $\mathcal{F}(t,\gamma_{ij})$ is a scalar function of the metric making $\cS$ invariant under 3-D diffeomorphisms. Indeed, the variation of such a functional wrt the metric will yield a covariantly conserved tensor which will thus satisfy (\ref{cov-cons}). In 3 dimensions all information about the spacetime curvature is contained in the Ricci tensor and $\mathcal{F}$ can be written as \cite{Croudace:1993yt,Stewart:1994wq}
\beq\label{gradient-exp}
\mathcal{F}=-2H(t)+J(t)R+L_1(t)R^2+L_2(t)R^{ij}R_{ij}+\ldots\,,
\eeq
a series in powers of the 3-D Ricci curvature involving an increasing number in gradients of $\gamma_{ij}$. $H(t)$ will turn out to be the Hubble rate. Note that under our assumptions this form is essentially unique. The Hamilton-Jacobi equation (\ref{H-J}) can now be solved separately for terms with different number of gradients. This gives for the time dependent coefficients in (\ref{gradient-exp})
\bea\label{H}
\frac{dH}{dt}+\frac{3}{2}H^2=0\,\\
\frac{dJ}{dt}+J H-\frac{1}{2}=0\,\\
\frac{dL_1}{dt}-L_1 H-\frac{3}{4}J^2=0\,\\
\frac{dL_2}{dt}-L_2 H+2J^2=0
\eea
e.t.c.
The solution for $H$ then determines all other functions:
\bea
&&H=\frac{2}{3t}\,,\\
&&J = \frac{3}{10}t\,,\\
&&L_1 = \frac{81}{2800}t^3\,,\quad L_2=-\frac{27}{350}t^3\,.
\eea

Given the above, the metric can obtained by solving equation (\ref{dotgamma}), which, up to terms with 4 spatial derivatives reads
\bea\label{dotgamma-4}
\frac{\partial\gamma_{ij}}{\partial t}&=&2H\gamma_{ij} + J(R\gamma_{ij}-4R_{ij})\nonumber\\
&&\hspace{-0.3cm}+L_1\left(3\gamma_{ij}R^2-8RR_{ij}+8R_{| ij}\right)\nonumber\\
&&\hspace{-0.3cm}+L_2\Big(3\gamma_{ij}R^{kl}R_{kl}-8R_{ik}R^k{}_j+\gamma_{ij}R^{|k}{}_{|k}\nonumber\\
&&\hspace{-0.3cm}+4R_i{}^k{}_{|jk}+4R_j{}^k{}_{|ik}-+4R_{ij}{}^{|k}{}_{|k}
\Big)+\ldots\,,
\eea
where the ellipsis denotes terms with 6 ore more spatial derivatives. This equation can now be solved iteratively. Setting an initial metric $k_{ij}$ at time $t_{\rm i}$ and writing
\beq\label{metric-expansion}
\gamma_{ij}=\gamma^{(0)}_{ij}+\gamma_{ij}^{(2)}+ \gamma_{ij}^{(4)}+\dots\,,
\eeq
we obtain at 0th order
\beq\label{metric-0}
\gamma_{ij}^{(0)}=\left(\frac{t}{ t_{\rm i}}\right)^{4/3}k_{ij}\,,
\eeq
at 1st order
\beq\label{metric-3}
\gamma_{ij}^{(1)}=\frac{9}{20}\left(\frac{t}{t_i}\right)^{2}t_i^2\left[\hat{R}k_{ij} - 4\hat{R}_{ij}\right]\,,
\eeq
and at second onder
\beq\label{metric-5}
\gamma_{ij}^{(2)}=\frac{81}{350}\left(\frac{t}{t_i}\right)^{8/3}t_i^{4}\left[\left(-4\hat{R}^{lm}\hat{R}_{lm}+\frac{5}{8}\hat{R}_{
;k}{}^{;k}+\frac{89}{32}\hat{R}^2\right)k_{ij} -10\hat{R}\hat{R}_{ij}
+17\hat{R}^l{}_i\hat{R}_{lj}-\frac{5}{2}\hat{R}_{ij;k}{}^k+\frac{5}{8}\hat{R}_{;ij}     \right]\,,
\eeq
where we have ignored terms that are subdominant when $t\gg t_{\rm i}$.\footnote{Such terms ensure that $\lim\limits_{t\rightarrow t_{\rm i}}\gamma_{ij}=k_{ij}$} An important point to note is that \emph{all hatted quantities are evaluated from the initial metric} $k_{ij}$: $\hat{R}=R[k_{ij}]$.

Before proceeding, it is worth estimating the range of validity of the above gradient series approximation. In solving (\ref{dotgamma-4}) we assumed that terms with more gradients are less important and that an iterative solution can be constructed. From (\ref{metric-0}) and (\ref{metric-3}) we can then estimate that the expansion should be valid for $t\lesssim t_{\rm con}$ where
\beq\label{t-con}
t_{\rm con}\simeq \mathcal{O}(1-3)\frac{1}{t_i^2\hat{R}^{3/2}}\,.
\eeq
The exact coefficient is of course dependent on the form of the initial 3-metric $k_{ij}$. After this time the contributions of successive terms to the metric containing more gradients are no longer perturbatively ordered and the iterative solution to (\ref{dotgamma-4}) breaks down. However, we will argue below that this is enough to determine the non-linear collapse of over-dense regions and goes some way in describing the expansion of under-dense regions into the non-linear regime.

Let us now apply the above to our universe. Assuming the standard inflationary initial conditions, the initial seed metric takes the form
\beq
k_{ij}=\left(\frac{t_{\rm i}}{t_0}\right)^{4/3}\delta_{ij}\left(1+\frac{10}{3}\Phi(\vc{x})\right)
\eeq
where $\Phi(\vc{x})$ is the primordial Newtonian potential and we have scaled the seed metric such that in the absence of perturbations the scale factor would be unity today. Note that the lowest order in the above expansion is simply the separate universe approximation where each spatial point scales as a FRW, CDM, universe with scale factor $a\propto t^{2/3}$.

In a flat, homogeneous, universe, $t_0$ (the age of the universe) can be determined from the expansion rate, $H_0=\dot{a}/a$, or density, $\rho_0=1/(6\pi G t_0^2)$, observed today. When there is backreaction, assigning $t_0$ is not straightforward. Ideally we wish to set $t_0$ such that when $t=t_0$ we obtain the \emph{average} expansion rate and \emph{average} density observed today. However, until performing the full numerical calculation we do not know what the resulting average values of these quantities will be. The problem is exacerbated by the transfer function used to obtain $\Phi(\vc{x})$ because it will also depend on the value of $t_0$. Therefore, altering $t_0$ does more than just rescale time.

We choose to define $t_0$ through the Hubble parameter of the \emph{background}, homogeneous, universe when it is extrapolated to $t=t_0$. For $H_0^{\rm back}=100 \,h\,{\rm km s}^{-1}\, \rm{Mpc}^{-1}$ we assign
\beq
t_0=\frac{2}{3 H_0}.
\eeq
Throughout, we take $h=0.4$. It is true that in the observed universe indications are that $h\simeq0.7$ \cite{arXiv:1001.4538}. However this result is derived assuming a $\Lambda$CDM model where backreaction is assumed to be negligible. Our background model is a flat CDM universe; therefore if we take $h=0.7$ the density of that background universe today would be much larger than what is observed. If the backreaction is a small effect then the current average density of the universe will also be too large. The choice $h=0.37$ corresponds to a background model that has the same density today as the best fit $\Lambda$CDM model \cite{arXiv:1001.4538}, albeit a different expansion rate. If the backreaction is small then the average density in the model today will be close to the observed average density. This choice is of course arbitrary and any value for $h$ in the range $(0.37-0.7)$ would be equally appropriate. We have tested values throughout this range. The qualitative nature of our results and thus the conclusions we draw from them do not change - in particular, while backreaction can be non-negligible in our model, it can never simultaneously reconcile $H_0^{\rm avg}$ and $\rho_0^{\rm avg}$ with observations. In future work, with $\Lambda$ included, or with backreaction large enough to mimic $\Lambda$, it would be interesting to scan through \emph{all} the relevant cosmological parameters, including $h$, to find the parameter set that best fit the observations.

With $k_{ij}$ as the initial metric, we have from (\ref{metric-0}) - (\ref{metric-5})
\beq \label{metric-1-phi}
\gamma_{ij}^{(0)}+\gamma_{ij}^{(1)}+\gamma_{ij}^{(2)}\simeq\left(\frac{t}{ t_0}\right)^{4/3}\left[\delta_{ij}+3\,\left(\frac{t}{t_0}\right)^{2/3}{t}_0^{2}\Phi_{,ij}+\left(\frac{t}{t_0}\right)^{4/3}t_0^{4}\hat{B}_{ij}\right]
\eeq
where
\beq\label{eq:Bdef}
\hat{B}_{ij}=\frac{9}{28}\left[19\Phi_{,il}\Phi^{,l}{}_{,j}-12\Phi_{,ij}\Phi^{,l}{
}_{,l}
+3\delta_{ij}\left((\Phi^{,l}{}_{,l})^2-\Phi_{,lm}\Phi^{,lm}\right)\right]
\eeq
Given these approximations, the local density of matter can be obtained
from
\beq\label{density}
\rho(t,\vc{x})=\frac{1}{6\pi G \, t^2}\frac{1}{\sqrt{{\rm Det} \left[\delta_{ij}
+3\left(\frac{t}{t_0}\right)^{2/3}{t}_0^{2}\Phi_{,ij}+\left(\frac{t}{t_0}\right)^{4/3}t_0^{4}\hat{B}_{ij}\right]}}
\eeq
Note that if this expression is expanded to linear order one recovers the result
of linear perturbation theory for the density contrast in the synchronous gauge. \emph{Equations (\ref{metric-1-phi}) and (\ref{density}) are the main results of the gradient expansion we will be using in this paper.}

The timescale for which the above approximation for the metric is accurate is estimated to be
\beq
\frac{t_{\rm con}}{t_0}\simeq 3.4\times\frac{H_0^3}{\left(\nabla^2\Phi\right)^{3/2}}\,,
\eeq
in accordance to (\ref{t-con}). We see that for describing an inhomogeneous universe which resembles ours for the whole of $t_0$, the metric can include perturbations down to about $k=0.3 h\, {\rm Mpc}^{-1}$. Of course, even shorter scales can be accurately described but for shorter times. It is then plausible that the gradient expansion can accurately follow the whole history of our model universe at least containing inhomogeneities that have recently entered the non-linear regime.

Unfortunately, to properly study backreaction we need to include perturbations over a broader range of scales and follow them for the whole of $t_0$. It might seem that the gradient expansion is of limited use for this and that all hope of studying backreaction in this framework is lost. However, as we will argue below, the situation is not completely irrecoverable. With certain modifications the above metric can still qualitatively capture the behaviour of collapsing structures and can also qualitatively capture the behaviour of regions of the universe that empty out and begin to expand faster than the background for the whole of $t_0$.

When regions of the universe collapse, we can see that the collapse time roughly equals $t_{\rm con}$ and thus for these regions and one can reliably follow the collapse with the above formalism. The particles in these regions gather in a volume that is effectively zero compared to the volume of the non-collapsed regions of the universe and once this has happened a collapsed region decouples form the rest of the universe. At these points, the precise volume or the full behaviour of the metric is irrelevant for calculating backreaction. To model the backreaction in the universe accurately, the metric needs only to correctly label which regions will collapse and capture the behaviour of the expanding regions. We will discuss ways to achieve this below.

To go some way towards addressing this issue we will now apply the above approximations to a small initial spherical perturbation and compare the result to the exact solution. This will give at least a qualitative understanding of the behavior of the truncated metric and allow us to develop some intuition before considering the problem of backreaction.

\subsection{The spherical case}\label{sec:sphere}

In order to better assess the accuracy of the series expansions (\ref{metric-expansion}) - (\ref{metric-5}) and obtain some intuition we now compare with a simple model: a spherical concentration of dust representing an over-(under-)density. For simplicity we will assume that the over-(under-)density is uniform and thus characterized by a positive (negative) constant spatial curvature. The exact spatial metric inside the sphere takes the FRW form
\beq
d\vc{x}^2=a(\tau)^2\left[\frac{dr^2}{1-\kappa r^2}+r^2d\Omega^2\right]
\eeq
where the dynamics of the scale factor $a(t)$ are given by the Friedmann-Lemaitre equation
\beq
\left(\frac{\dot{a}}{a}\right)^2=\frac{8\pi G}{3}\frac{f_0}{a^3}-\frac{\kappa}{a^2}
\eeq
whose solutions can be written in parametric form
\bea
a(u)=\frac{8\pi G}{6}\frac{f_0}{|\kappa|}\left(\cosh u-1\right)\,,\quad \sqrt{|\kappa|}\tau(u)=\frac{8\pi G}{6}\frac{f_0}{|\kappa|}\left(\sinh u-u\right)\,,\quad (\kappa<0)\\
a(u)=\frac{8\pi G}{6}\frac{f_0}{\kappa}\left(1-\cos u\right)\,,\quad \sqrt{|\kappa|}\tau(u)=\frac{8\pi G}{6}\frac{f_0}{\kappa}\left(u-\sin u\right)\,,\quad (\kappa>0)
\eea
On the other hand, applying formulae (\ref{metric-expansion}) - (\ref{metric-5}) we obtain as approximations for the scale factor
\beq\label{approx-sph}
a(\tau)=\left(\frac{\tau}{\tau_0}\right)^{{2}/{3}}\sqrt{1 \mp \frac{9}{10}\, |\kappa|\,  \tau_0^{2} \left(\frac{\tau}{\tau_0}\right)^{2/3}+ \frac{81}{350} \frac{1}{8}\, \kappa^2\, \tau_0^{4} \left(\frac{\tau}{\tau_0}\right)^{4/3} + \ldots}
\eeq
where
\beq
\tau_0=\frac{1}{\sqrt{6\pi G f_0 }}.
\eeq

\begin{figure}[t]
\includegraphics[scale=0.7]{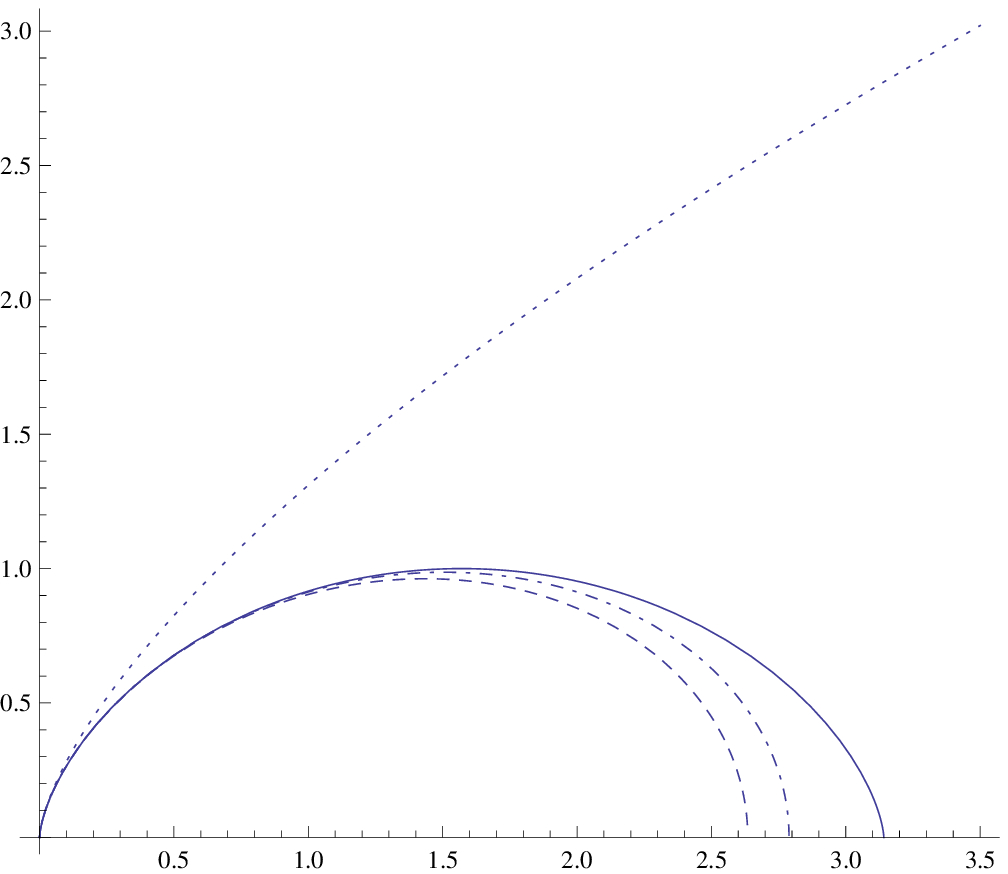}
\includegraphics[scale=0.7]{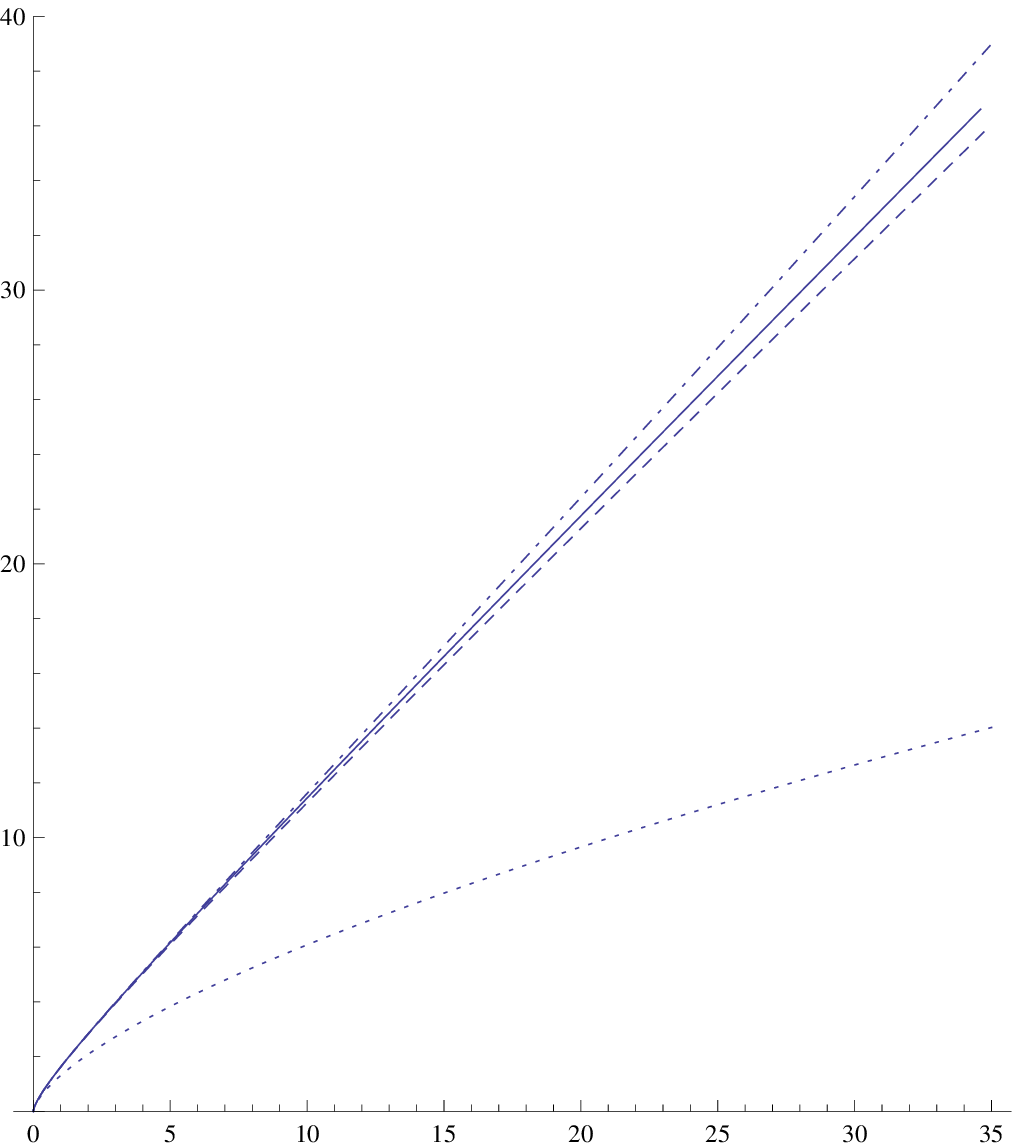}
\caption{\label{spherical} The radius for the $r=0$ shell of a spherically symmetric ball of dust with uniform density of positive (left) and negative (right) 3-curvature as a function of $\sqrt{|\kappa|}\tau$. We have set $8\pi G=3$ and $f_0/\kappa=1$. Larger values of $f_0/\kappa$, representing smaller curvatures, simply scale the above graphs to longer times and larger scale factors. The solid line is the exact solution, the dotted line is the 0th order approximation for the metric (flat FRW or "separate universe approximation"), the dashed line is the 1st order metric (with 2 spatial derivatives) and the dash-dotted line is the 2nd order metric (with 4 spatial derivatives). For positive curvature the addition of successive terms improves the convergence. For negative curvature, successive terms eventually deviate from the exact solution as the radius of convergence for the series is exceeded. However, a good description is obtained for relatively large values of $\sqrt{|\kappa|}\tau$.}
\end{figure}

The comparison of the approximation to the exact solution is shown in figure \ref{spherical}. It is immediately apparent that the first term in the series already goes beyond linear theory in capturing the eventual collapse of a positively curved region (an initial overdensity in a perturbed FRW model). Additional terms improve the convergence to the exact behavior and predict the collapse time more accurately. This is in agreement with the expectations of the previous section that the series converges to the exact solution for times comparable to the collapse time.

For regions with negative curvature (modeling initial underdensities) the first term is pretty close to the exact solution, reproducing the asymptotic $a\propto \tau$ behavior of an eventually empty region which is simply Minkowski space written in expanding coordinates. Although the additional terms improve the convergence initially, it is clear that at late times these terms deviate from the asymptotic behavior. Close examination reveals that this occurs at around the collapse time $\tau_{\rm col}\sim \frac{f_0}{\kappa^{3/2}}$ of the corresponding configuration with positive curvature, as expected in the previous section. However, the underdense regions are still reasonably well described up to times much longer than the collapse times of regions with corresponding positive curvatures: from (\ref{approx-sph}) we see that at $\tau\sim 10 \tau_{\rm col}$ the solution deviates from the exact one by a few percent.

The success of the gradient expansion for this simple case gives us confidence that our metric, eq.\eqref{metric-1-phi} can be used to model, at least \emph{qualitatively}, the salient features of inhomogeneities in our universe that are relevant for backreaction: the collapse of overdense regions and the expansion of underdense ones \emph{significantly beyond the linear
regime and in a manner more realistic than the models used so far.}

\subsection{Potential complications for application to backreaction}\label{sec:probs1}

The previous subsection has indicated that the metric obtained form the gradient expansion can be a good model for studying backreaction. However, there are potential pitfalls when applied to more realistic distributions of matter that must be tackled before applying it to this problem. Firstly, in the real universe the metric at each point will not be spherically symmetric. Most importantly, there will be off-diagonal elements in the metric that are not present in the simple case of spherical symmetry. As we will see in section \ref{sec:sphregions} these terms can have a significant effect on the behaviour of the truncated metric. In section \ref{sec:sphregions} we attempt to address this complication.

Also, in the real universe, neighbouring regions interact. An initially underdense expanding region will not continue to expand as if it were isolated. Eventually it will expand far enough to hit other structures, which will necessarily affect the expansion. One would expect that as long as the series is convergent and with a sufficient number of terms retained, the metric in the gradient expansion should capture this behavior. However, for practical reasons we need to truncate the metric at a low order in gradients and eventually exceed the times for which the series converges. The fact that a low order truncated metric correctly describes the behavior in an isolated universe, as shown above, is not proof that it will also work well when there \emph{are} interactions between neighboring regions. As we will see in section \ref{sec:fixvoid}, the expected behaviour of expanding regions in the real universe, when the regions interact, is not perfectly captured by the truncated metric. In section \ref{sec:fixvoid} we also attempt to address this complication.

Finally, the synchronous gauge allows coordinate singularities where the determinant of the metric becomes $0$ and the density apparently diverges. These occur when the trajectories of particles, and consequently the coordinate lines they define, cross and caustics form. It is difficult to distinguish such a coordinate singularity from a region that is actually collapsing. One possible method is to calculate the Ricci scalar of the full 4-metric. At a coordinate singularity, the 4-Ricci scalar will not grow large, whereas in a true collapsing region it should at least grow larger than the background. However, even if regions could be correctly identified as coordinate singularities it would still be difficult to deal with them. Arguably, one could transfer into a more stable coordinate system, use that to travel through the singularity and then return to a new synchronous coordinate system. Remember, however that at these coordinate singularities particles are crossing each other's paths. There will inevitably be significant \emph{non-gravitational} effects at these points. Even for weakly interacting dark matter particles, the single velocity fluid description will break down at these points. These effects are difficult to model, but they will certainly inhibit the expansion of the volume occupied by the crossing particles.

To properly compare this model to the observed universe, this issue will need to be addressed more precisely.\footnote{A useful approach for future improvements might be an adaptation in this setting of the ``adhesion approximation'' described in \cite{Buchert:2005xj}.} However, for our purposes, because we expect the dynamics of collapsed regions and regions with shell-crossings to be similar,\footnote{Even if their expansion is inhibited for different reasons.} we have chosen to label every region where the density diverges as a collapsed region. We expect that since any backreaction will be dominated by the expanding regions the accurate modeling of the contracting ones will not be critical. We explain in section \ref{sec:fixvoid} how we deal with the evolution of contracting regions in our model.

\subsection{Backreaction in the synchronous gauge}

Let us now briefly recall the definitions for backreaction and apply them to the synchronous gauge. Consider a fluid filled universe with $u^\mu$ the four-velocity of the fluid.
The local expansion of the fluid lines is given by
\beq
\theta=u^\mu{}_{;\mu}
\eeq
while the stress tensor reads
\beq
\sigma_{\mu\nu}=\frac{1}{2}\left(u_{\nu;\mu}+u_{\mu;\nu}\right)-\frac{1}{3}
\left(g_{\mu\nu}+u_\mu u_\nu\right)\theta.
\eeq
When one averages aver a domain D the average scale factor
\beq\label{eq:defaD}
a_D(t)^3\equiv \int_D d^3 x \sqrt{\gamma(t,\vc{x})}
\eeq
obeys the Buchert acceleration equation \cite{Buchert:1999er}
\beq\label{eq:Qfroma}
3\frac{\ddot{a}_D}{a_D}=-4\pi G \langle\rho\rangle_D 
+ Q_D
\eeq
where
\beq
Q_D=\frac{2}{3}
\left(\langle\theta^2\rangle_D-\langle\theta\rangle_D^2\right)-\langle\sigma_{\mu\nu
} \sigma^{\mu\nu} \rangle_D
\eeq
is the backreaction term. The brackets denote a spatial average
\beq
\langle S\rangle_D = \frac{\int_D d^3x \,\sqrt{\gamma}\, S}{\int_D d^3x
\, \sqrt{\gamma }}\,.
\eeq
The above equation differs from the standard FRW one by the $Q$ term, making it
obvious that inhomogeneities affect the evolution of the average scale factor.
In the synchronous gauge it reads
\beq\label{eq:Qdef}
Q_D=\frac{1}{4}\langle\left(\gamma^{ij}\dot{\gamma}_{ij}\right)^2\rangle_D-\frac{1}{
6}\langle\gamma^{ij}\dot{\gamma}_{ij}\rangle_D^2
-\frac{1}{4}\langle
\gamma^{ki}\dot{\gamma}_{ij}\gamma^{jl}\dot{\gamma}_{lk}\rangle_D
\eeq
As is implied by the notation above, the average quantities depend on the averaging domain and therefore the scale on which the averaging is performed. From now on we will mostly drop the the explicit reference to D but the inherent scale dependence for the quantities we refer to should be kept in mind.

We can now use the expressions for the metric described above to estimate the
amount of backreaction in our universe. We will directly evaluate the behavior
of the average scale factor and cross-check with an explicit determination of
$Q_D$.  The evaluation requires the determinant of the metric and its inverse
$\gamma^{ij}$. Even though we have an explicit form for $\gamma_{ij}$ in eq.\eqref{metric-1-phi}, writing an explicit expression for
the determinant and the inverse is not convenient. Note that we cannot expand in
a series since we are interested in the regime where the corrections from the
gradient terms become order 1. It is however straightforward to approach the
problem numerically as we describe in the following section.

\section{Numerical implementation}\label{sec:numimp}
To calculate the backreaction numerically, we set up an $N^3$ grid. We assign
the initial gravitational potential $\Phi(\vc{x})$ on this grid to be a gaussian
random field consistent with the known cosmology. Then, the evaluation of two
spatial derivatives provides the matrices $\Phi_{,ij}$ and $\hat{B}_{ij}$ in (\ref{metric-1-phi}). Once these are obtained, expression (\ref{metric-1-phi}) defines at each point on the grid a $3\times 3$ symmetric matrix which provides an explicit time dependent model for the 3-metric. It is then straightforward to
find the determinant ${\gamma}$ at each point and hence the average scale factor
$a_D(t)$ by integrating over the points of the grid. Furthermore, the inverse
$\gamma^{ij}$ can also be calculated at each point and hence $Q$ can be
obtained.

\subsection{Setting up the initial gravitational potential $\Phi(\vc{x})$}
The initial metric perturbation is (approximately and sufficiently for this
work) a gaussian random field. Since we'll be working in a box of size $L$ we
use a fourier series
\beq
\Phi(\vc{x})=\sum\limits_{\vc{k}}\Phi_\vc{k} \, e^{i\vc{k}\cdot\vc{x}}
\eeq
with periodic boundary conditions such that $\vc{k}=\frac{2\pi}{L}
\left(l,m,n\right)$, with $l,m,n=0,\pm 1, \pm 2,\ldots$. Reality of the field
implies $\Phi_{\vc{k}}^\star=\Phi_{-\vc{k}}$. Writing
$\Phi_{\vc{k}}=R_\vc{k}+iI_\vc{k}$, the real and imaginary parts $R_\vc{k}$ and
$I_\vc{k}$ are gaussianly distributed random variables. The \emph{correlation
function} in a specific realization of the random field is given by
\beq
\xi(\vc{r}) \equiv \langle \Phi(\vc{x})\Phi(\vc{x}+\vc{r})\rangle_V
\eeq
where $\langle \ldots \rangle_V$ is a \emph{volume average}. We then have
\beq
\xi(\vc{r})=\sum\limits_{\vc{k}}\left(R_{\vc{k}}^2+I_{\vc{k}}^2\right)e^{i\vc{k}\cdot
\vc{r}}
\eeq
The variance of $R_{\vc{k}}$ and $I_{\vc{k}}$ will be $\langle
R_{\vc{k}}^2\rangle_E=\langle I_{\vc{k}}^2\rangle_E=\frac{1}{2}\sigma_{k}^2$,
depending only on the magnitude of the wavevector. The \emph{ensemble average}
of the correlation function then reads
\beq
\langle \xi(\vc{r}) \rangle_E = \sum\limits_{\vc{k}}  \sigma_k^2
\,e^{i\vc{k}\cdot\vc{r}}
\eeq
Given a large number of modes and the ergodic assumption, the correlation
function measured in our universe $\xi_M(r)$ can be interpreted as $\langle
\xi(\vc{r}) \rangle_E$ and thus $\sigma_k$ is determined by the measured
power-spectrum.  If our box is large enough such that $\sum\limits_{\vc{k}}
\rightarrow \frac{V}{(2\pi)^3}\int d^3\vc{k}$ we have
\beq
\sigma_k^2 =
\frac{9}{25}\frac{2\pi^2}{k^3}\frac{\Delta^2_\mc{R}}{V}\left(\frac{k}{k_\mathrm{
piv}}\right)^{n_s-1} T(k)^2
\eeq
where $T(k)$ is the transfer function that evolves the primordial gravitational potential through the radiation dominated era. Throughout, we use the transfer
function of \cite{Bardeen:1985tr} with the baryonic correction presented in \cite{Sugiyama:1994ed}. For the primordial
spectrum we take, at the pivot scale $k_\mathrm{piv}=0.002$ Mpc, the spectral index to be $n_s=0.963$ and the amplitude $\Delta^2_\mc{R}=2.42\times 10^{-9}$ as measured by WMAP \cite{arXiv:1001.4538}. As inputs to the transfer function we also take $h=0.4$ and $\Omega_b=0.17$.\footnote{Each of these are quantities of the background metric at $t=t_0$, not the full backreacting CDM universe.} We choose this value of $\Omega_b$ because it preserves the ratio between $\Omega_b$ and $\Omega_m$ observed in the real universe.

To create a given realization one has to draw $R_\vc{k}$ and $I_\vc{k}$ of each
fourier mode in the box from a random process with a gaussian probability
distribution
\beq
P(X_\vc{k})dX_\vc{k}=\frac{\sqrt{2}}{\sigma_k\sqrt{2\pi}}\,\exp\left[{-\frac{
X_\vc{k}^2}{\sigma_{k}^2}}\right]\,.
\eeq
The probability $dP$ to obtain a given value of the complex $\Phi_\vc{k}$ is
\beq
dP=P(R_\vc{k})P(I_\vc{k})dR_\vc{k}dI_\vc{k}.
\eeq
Changing variables to $\Phi_\vc{k}=A_\vc{k}\,e^{i\theta_\vc{k}}$ we get
\beq
dP=\frac{2
A_\vc{k}}{\sigma_k^2}\,\exp\left[{-\frac{A_\vc{k}^2}{\sigma_{k}^2}}\right]
dA_\vc{k} \, \frac{d\theta_\vc{k}}{2\pi}\,.
\eeq
We see that the amplitude is described by a \emph{Rayleigh distribution} and the
phase is uniformly distributed in the range $[0,2\pi)$. Using this, we can also
set up a realization of the gravitational potential and its derivatives from
\beq
\Phi(\vc{x})=2\sum\limits_{\vc{k}\in {\rm uhs}}
A_\vc{k}\cos\left(\vc{k}\cdot\vc{x}+\theta_\vc{k}\right)\,,
\eeq
and
\beq\label{eq:ddphidef}
\Phi_{,ij}(\vc{x})=-2\sum\limits_{\vc{k}\in {\rm uhs}}
k_i k_j A_\vc{k}\cos\left(\vc{k}\cdot\vc{x}+\theta_\vc{k}\right)\,,
\eeq
This provides all the necessary information
for the evaluation of the metric at each point \emph{for all times} using
eq.\eqref{metric-1-phi}.

Note that, for the derivative $\Phi_{,ij}(\vc{x})$, the sum is only very weakly convergent. It is instructive to count, in the continuum and large $k$ limit, the factors of $k$ involved in the variance of $\Phi_{,ij}(\vc{x})$. From $\sigma_k^2$ there are factors of $k^{-3}$, $T(k)^2$ and $k^{n_s-1}$, from the second order derivative there are two factors of $k^2$, and from the integral measure there is a factor of $k^3$. In the large $k$ limit, $T(k)\rightarrow \ln(k)/k^2$. Therefore, the value of the partial sum, up to a value of $k$, scales as $\ln(k)^2\times k^{n_s-1}$. While in principle the red-tilted power law $k^{n_s-1}$ will eventually fall more quickly than the growing $\ln(k)^2$ term, this will only happen at very small scales (because $\left| n_s-1 \right|\ll1$). This becomes important in the following section.

\subsection{Dealing with the complications and presenting the backreaction}

Ideally we would like to use eq.\eqref{eq:ddphidef} to generate $\Phi_{,ij}(\vc{x})$, then substitute this into eq.\eqref{metric-1-phi} and finally use eqs.\eqref{eq:defaD} and \eqref{eq:Qdef} to evaluate the average scale factor $a_D$ and the magnitude of the backreaction, $Q$, in our choice of CDM universe. However there are a number of complications that arise before this can be straightforwardly achieved. Some of these were flagposted in section \ref{sec:probs1} and relate to the truncation of the metric or the extrapolation to long times, the use of the synchronous gauge and the almost divergent nature of the sum in eq.\eqref{eq:ddphidef}.

The gradient formalism presented so far involves no assumptions about the universe except that it is dominated by dust, was once flat and homogeneous and is well described by general relativity. Every new feature that we must now add to deal with the complications arising adds extra assumptions about the universe. Therefore we are not able to calculate the backreaction from all scales in a dust filled universe with certainty. However, we can provide a plausible \emph{candidate model} to describe the backreaction. The backreaction in the CDM version of this model is not large enough to mimic $\Lambda$ but would be large enough to be measurable. If this model were extended to include $\Lambda$ it could be tested against observations. It would be interesting to see if it provides a better fit than standard $\Lambda$CDM for a wider set of data.

We discuss in section \ref{sec:extensions} various ways in which this model can be improved and refined beyond just the obvious inclusion of $\Lambda$. Nevertheless, we argue that even in its current form, the model resembles the actual universe more closely than earlier models in the literature that deal with backreaction. Once the initial conditions are set up, each grid point evolves independently from the others. However, we show in section \ref{sec:univsli} that there is a depth of structure in the evolving box of grid points. This is because the initial conditions of the metric at each gridpoint are not independent.

In the rest of this section we proceed to discuss the complications encountered when applying the gradient formalism to the problem, our methods for dealing with them and the corresponding results for the backreaction.

\subsubsection{The almost UV divergence}\label{sec:byeUV}

The sum in eq.\eqref{eq:ddphidef} converges as $\sim \ln(k)/k^{(1-n_s)/2}$. However $\ln(k)/k^{(1-n_s)/2}$ does not begin to decrease until $k\gg 1$. It is highly likely that other natural cutoffs such as the free-streaming length of dark matter will regulate this sum long before it converges. Therefore it is not tractable (or even physically motivated) to calculate the value of this sum in the limit $k\rightarrow \infty$. More importantly however, structure growth at any given scale will freeze out after some time. No kpc sized scales are still evolving gravitationally today. They have either collapsed into bound structures or are part of much larger (in synchronous coordinates) expanding regions.

There are also practical limitations. Firstly, a box of $N^3$ grid points with a grid spacing that corresponds to the natural cutoff of eq.\eqref{eq:ddphidef} will either need to be very small, or have very many points in it. We obviously cannot have an arbitrarily large number of grid-points without an arbitrarily powerful computer. If instead we choose to examine a small box then we lose information about correlations over distances greater than the box size. The model therefore loses the ability to calculate the backreaction that arises from structure growth over those scales. Secondly, we saw above that under-densities on short scales cannot be accurately followed by the gradient expansion for a arbitrarily long times unless extra assumptions are introduced made about their late time behaviour. Thus, even if we could include all scales from the very long to the very short in eq.\eqref{eq:ddphidef} it would be unreasonable to expect that our approximations would make sense.

We are interested in the scales that are evolving most significantly \emph{today}. Therefore, we choose to examine boxes where the grid-spacing corresponds to these scales. The near divergent nature of eq.\eqref{eq:ddphidef} means that it will then be these scales that dominate the value of $\Phi_{,ij}(\vc{x})$ and thus the behaviour of the evolving metric. By varying the number of grid points we then vary the size of the box. For a box with grid spacing $R$ and number of points $N_x$ its size is $L=R N_x$. We then sum eq.\eqref{eq:ddphidef} from the value $k_{\rm min}=(2\pi)/L$ up to the value $k_{\rm max}=(2\pi)/R$.

Although this is necessary to have a working model, it does have the drawback that all evolution of the universe at smaller scales is removed from the calculation. Our model sees the universe as a collection of synchronous ``blobs''. These blobs expand and contract, but any internal structure of each blob is ignored. Similarly, as argued above, any backreaction at much larger scales than the grid spacing is not seen because of the finite box size. We cannot therefore capture \emph{all} the backreaction in the universe over \emph{all} times with our model. What we hopefully do capture is the backreaction that occurs due to the scales we are examining. In our plots we choose to focus on approximately the scales that have the largest backreaction from $t\simeq t_0/2$ until $t=t_0$. This corresponds to $R=1\, {\rm Mpc}/h$. In figure \ref{fig:Omeg_X} we show the backreaction over a range of grid spacings. In section \ref{sec:extensions} we discuss possible methods to construct a model in the future that accounts for backreaction over all scales.

\subsubsection{Fixing the metric}\label{sec:fixvoid}

\paragraph{Over-densities:} If we let the metric evolve without any alterations then in all of the collapsing regions (i.e. grid points with increasing density) the determinant of the metric will eventually change sign. When this happens, both the local density and local volume will become imaginary. Clearly this is both unwanted and unphysical. In the real universe when a region collapses it virialises and its growth is subsequently frozen. We choose to model this by freezing the metric of any grid point when the ratio between the density at this grid point and the background density has crossed a certain threshold, $\rho_{\rm rat}$.\footnote{See section \ref{sec:probs1} for a discussion concerning the potential misidentification of coordinate singularities as collapsed regions.} This corresponds to a condition on the determinant of the metric, $\gamma(t,{\bf x})$ at each grid point of
\beq\label{eq:collapse}
\gamma(t,{\bf x})<\frac{\left(\frac{t}{t_0}\right)^4}{\rho_{\rm rat}^2}.
\eeq
For every grid-point, ${\bf x}^*$, where this condition is satisfied at some time $t^*({\bf x}^*)$, we define $\gamma_{ij}(t,{\bf x}^*)=\gamma_{ij}(t^*,{\bf x}^*)$ for all $t>t^*$. By extension, for the purposes of calculating $Q$, we also define $\dot{\gamma}_{ij}(t,{\bf x}^*)=0$ for all $t>t^*$. We find that once $\rho_{\rm rat}$ is set to a sufficiently large value the precise value of $\rho_{\rm rat}$ has little effect on the results for $Q$ and $a_D$. This is not surprising given that, independent of the value of the threshold, the asymptotic state for any of these grid-points is a point of zero volume relative to the volume of all of the expanding grid-points.

\paragraph{Under-densities:} We showed in section \ref{sec:sphere} that the truncated metric describes the evolution of an isolated, under-dense, spherical region well. Unfortunately, the universe is not made up of a collection of truly isolated points. If the truncated metric is left unaltered then expanding regions will continue to expand faster than the background, indefinitely. When keeping just one gradient term in the truncated metric, the asymptotic state of each under-dense point is equivalent to an empty, open universe. When keeping two terms the asymptotic state expands even faster. This behaviour results from the series not converging anymore on such timescales. 
In the real universe the expanding regions will eventually hit walls and filaments that will impede (or perhaps even promote) their expansion. At this point the region corresponding to a given grid-point will continue evolving as a part of a bigger over-all region. That is, larger scales will have become relevant for studying structure growth.

We do not know what will happen to each grid-point when these larger scales become relevant. The only way to know is 
to ``zoom out'' and recalculate the universe in a larger box, with a larger grid-spacing. Unfortunately, different under-dense grid-points will expand at different rates. Therefore, they will expand into neighbouring regions at different times. So, we cannot just evolve the metric until larger scales become relevant and then stop. Some regions will reach this point much earlier than others and the most extreme regions will reach this point very early on. At the very least, we need a way to deal with the most extreme regions, but ideally we also want to know at all times what magnitude of backreaction has been generated by the scale being examined. This includes times long after other scales have also become relevant.

We attempt to deal with the expanding regions by taking any gridpoint that has over-expanded and fixing it to evolve with the background expansion. We tag a grid-point as over-expanded when the ratio of its volume to the comoving background volume at that gridpoint has exceeded a certain threshold, $g_{\rm rat}$. That is, when
\beq\label{eq:expand}
\gamma(t,{\bf x})>g_{\rm rat}^2 \left(\frac{t}{t_0}\right)^4.
\eeq
For every gridpoint, ${\bf x}^*$, that satisfies this condition at some time, $t^*({\bf x}^*)$, we fix the metric to have the following behaviour
\begin{eqnarray}\label{eq:voidfix}
\gamma_{ii}&=&\gamma^{1/3} \left(\frac{t}{t^*}\right)^{4/3} \nonumber \\
\gamma_{ij}&=&0 \, , \,\, i\neq j \nonumber \\
\dot{\gamma}_{ii}&=&\frac{4}{3t}\gamma_{ii} \nonumber \\
\dot{\gamma}_{ij}&=&0 \, , \,\, i\neq j
\end{eqnarray}
for all $t\gg t^*$. Note, $\gamma=\gamma(t^*,{\bf x}^*)$. If we impose this condition immediately at $t=t^*$ the discontinuous change in $\dot{\gamma}$ causes numerical artifacts to appear in $Q$. Therefore we actually impose this transition smoothly in a manner described in appendix \ref{app:smooth}, with eq.\eqref{eq:voidfix} only the asymptotic final state of any over-expanded gridpoint. This transition can be made arbitrarily fast; however see appendix \ref{app:smooth} for a discussion. Unlike the situation for collapsed regions, our results are highly sensitive to the value of $g_{\rm rat}$. This is also not surprising given that it is the expanding regions that come to dominate the volume of the universe.

The motivation for this particular method of fixing expanding regions is that once a region has crossed this threshold, any backreaction that occurs is due to evolution on larger scales. Therefore the backreaction caused at this gridpoint by the scale of interest is, arguably, zero. By fixing the grid point to evolve with the background we force there to be no additional backreaction from that region.

Note that even if the complete solution for the metric could be obtained it would still not be an entirely realistic description of the non-linear regime. Indeed the contraction of overdensities is eventually halted by further baryonic physics and structures virialize (unless they are dense enough to form black holes) while underdensities are rarefied up to the point when the shells surrounding them cannot expand further due to the interaction with surrounding overdensities. Although apparently artificial, this treatment of non-linear scales should in principle capture the effect of the actual baryonic physics preventing gravitational collapse and leading to virialized structures which decouple from the background evolution.

\subsubsection{Two definitions of $Q$ and $a$}\label{sec:aQdefs}

We want to calculate the values of $a_D$ and $Q$ in our model universe. The definitions for $Q$ provided in eqs.\eqref{eq:Qfroma} and \eqref{eq:Qdef} assume a universe composed only of dust. Ostensibly, this is true for the model. However, as soon as we 
fix expanding regions to the background by the ``hand of God'' there is a risk that the metric will behave in a way that no longer describes a dust only universe. If the universe behaves in a non-dusty manner then we expect that the $Q$ calculated from the two equations will differ. This is because there will be an effective pressure term that \emph{should} be present in eq.\eqref{eq:Qfroma}.

We do not want our somewhat ad hoc fixing of the expanding regions to be interpreted as due to the effects of backreaction. Therefore, we calculate $Q$ in two manners.\footnote{For a related Newtonian analysis see \cite{Buchert:1999pq}} The first method, which we still label as $Q$, we calculate directly from eq.\eqref{eq:Qdef}. The second method involves numerically differentiating $a_D(t)$ twice to obtain $\ddot{a}_D(t)$ and using this to define an effective $Q(a_D)$ from eq.\eqref{eq:Qfroma}. Correspondingly, we also calculate $a_D$ in two ways. The first is directly from the definition, eq.\eqref{eq:defaD}. The second method involves numerically integrating eq.\eqref{eq:Qfroma} using $Q$ as calculated from eq.\eqref{eq:Qdef}.

It is the calculation of $Q(a_D)$ that is most responsible for our need to impose the fix of section \ref{sec:fixvoid} in the smooth manner described in appendix \ref{app:smooth}. To calculate $Q(a_D)$ we require $\ddot{a}_D$. If the evolution of $\gamma$ at any point is not smooth, then the evolution of $a_D$ is not either. Thus, $\dot{a}_D$ is not continuous and $\ddot{a}_D$ is not well-defined. Even with the smooth fix applied to the expanding regions $Q(a_D)$ exhibits a large degree of numerical noise. In figures
\ref{fig:Qsphere} and \ref{fig:Qsquare} we first smooth $a_D$ over a small range of time before taking the numerical derivative to obtain $\dot{a}_D$ and $\ddot{a}_D$.

It could be argued that the best method to fix $\gamma_{ij}$ in the expanding regions is the one for which both methods of $Q$ and both methods of $a_D$ align most closely.

\subsubsection{Backreaction with artificial local isotropy}\label{sec:sphregions}

The determinant of the metric tensor is given by
\beq\label{eq:detform}
\gamma=\gamma_{11}\gamma_{22}\gamma_{33}+2\gamma_{12}\gamma_{23}\gamma_{13}-\gamma_{12}^2\gamma_{33}-\gamma_{23}^2\gamma_{11}-\gamma_{13}^2\gamma_{22}\, .
\eeq
To first order in the expansion, the metric tensor itself is given by (see eq.\eqref{metric-1-phi})
\begin{eqnarray}
\gamma_{ii}&=&\left(\frac{t}{t_0}\right)^{4/3} \left(1+3t^{2/3}t_0^{4/3}\Phi_{,ii}\right) \nonumber \\
\gamma_{ij}&=&\left(\frac{t}{t_0}\right)^{4/3}\left(3t^{2/3}t_0^{4/3}\Phi_{,ij}\right) \, , \,\, i\neq j\, .
\end{eqnarray}
$\Phi_{,ij}$ is distributed symmetrically around zero. Therefore, if we ignore correlations, the first term in $\gamma$ is symmetrically distributed around $(t/t_0)^4$. The second term is symmetrically distributed around zero. However, $\Phi^2_{,ij}$ is always positive. Therefore, the third, fourth and fifth terms in $\gamma$ will not be symmetrically distributed around zero. The net result is that $\gamma$ is not symmetrically distributed around $(t/t_0)^4$, but is strongly skewed towards $\gamma<(t/t_0)^4$.

If we naively apply the formalism as it has been developed up to this point, then our model universe will consist almost entirely of collapsed regions. It is tempting to interpret this result as physically motivated; however it is not. It is an artifact of the truncation of the metric. While we have been consistent in keeping the same number of gradient terms at each order of the expansion in the \emph{metric}, this is not true for the metric's determinant. Even when we include just one gradient term in the metric, the metric's determinant has terms up to cubic order in the gradient. It might then be tempting to also truncate the determinant at a certain order of the gradients; however this makes the model itself inconsistent. We may have truncated the metric, but the terms in the metric are not perturbatively ordered; therefore removing terms from the metric's determinant would mean that our model no longer satisfies general relativity. That is, we would not be using the correct volume element for the model's metric tensor when calculating $Q$ and $a_D$.

We solve this, the last of the model's complications, in two ways.\footnote{The second method is introduced in section \ref{sec:squared}.} In this section, we consider a simplification of the metric such that $\Phi_{,ii}=\Phi_{,11}$ and $\Phi_{,ij}=0$ for all $i \neq j$.\footnote{Note that we do not leave $\Phi_{,22}$ and $\Phi_{,33}$ as their true values because the non-zero correlation between $\Phi_{,ii}$ and $\Phi_{,jj}$ also skews $\gamma$ towards collapsed regions.} This amounts to considering every grid-point in our model as a spherically symmetric region. While this choice of initial condition is less well motivated than using the full gradient $\Phi_{,ij}$, once the initial condition is set up it behaves entirely consistently. Moreover, it provides a good point of comparison. We know from section \ref{sec:sphere} that the truncated metric describes an isolated spherical region well. Therefore, up until the point when larger scales should become relevant, we can be confident that the metric will describe the evolution of these initial conditions accurately. Although much of the information about the correlations between different grid-points is removed because of this simplification, we show in section \ref{sec:univsli} that not all of it is.

Note that even in this simplified model, the distribution of $\Phi_{,11}$ itself matches that expected from a near scale-invariant, Gaussian, power-law primordial spectrum. Therefore, while information about correlations between points is lost, the points themselves should have the correct distribution.

\begin{figure}[t]
\includegraphics[scale=0.4]{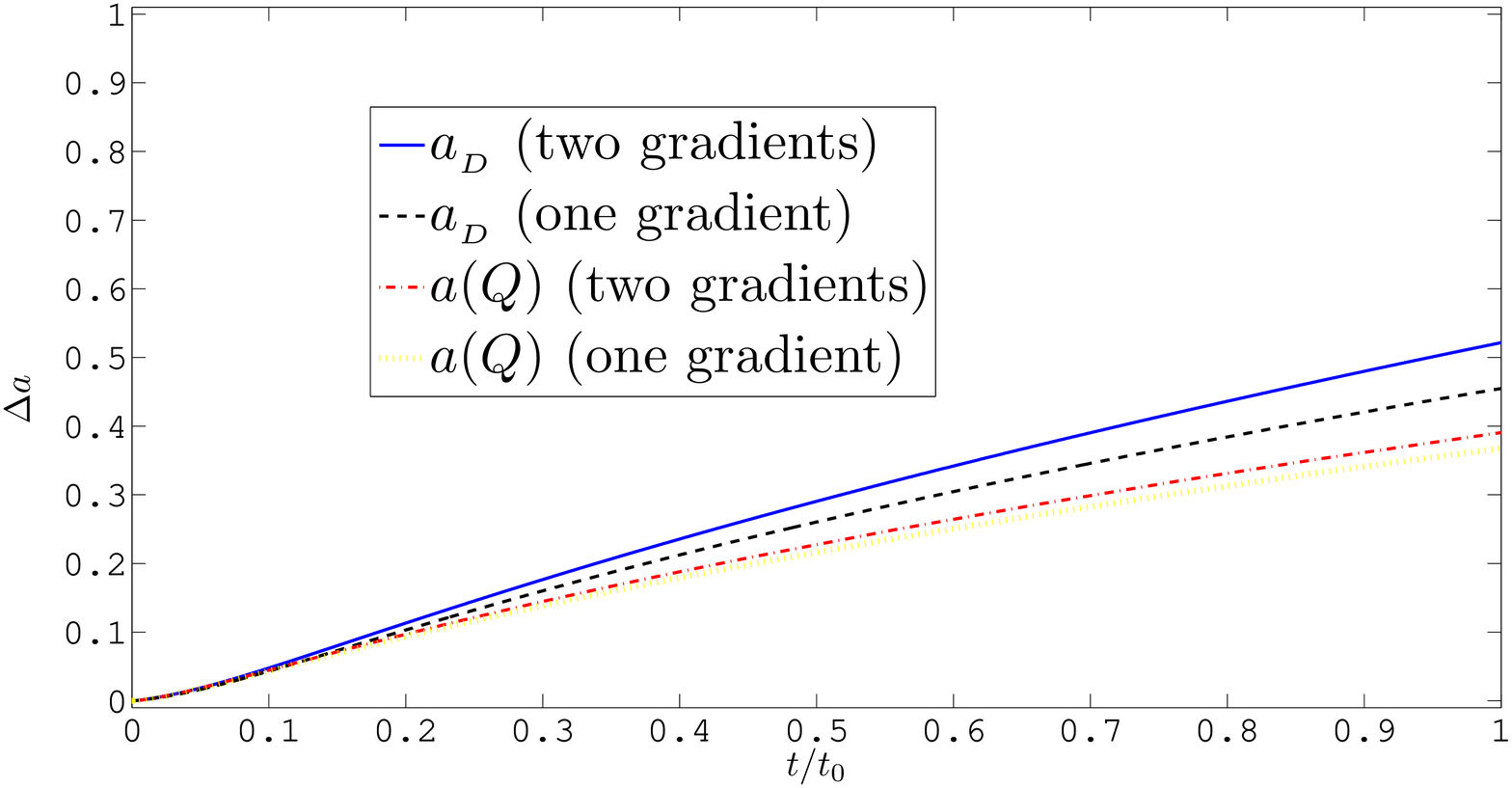} \\
\includegraphics[scale=0.2]{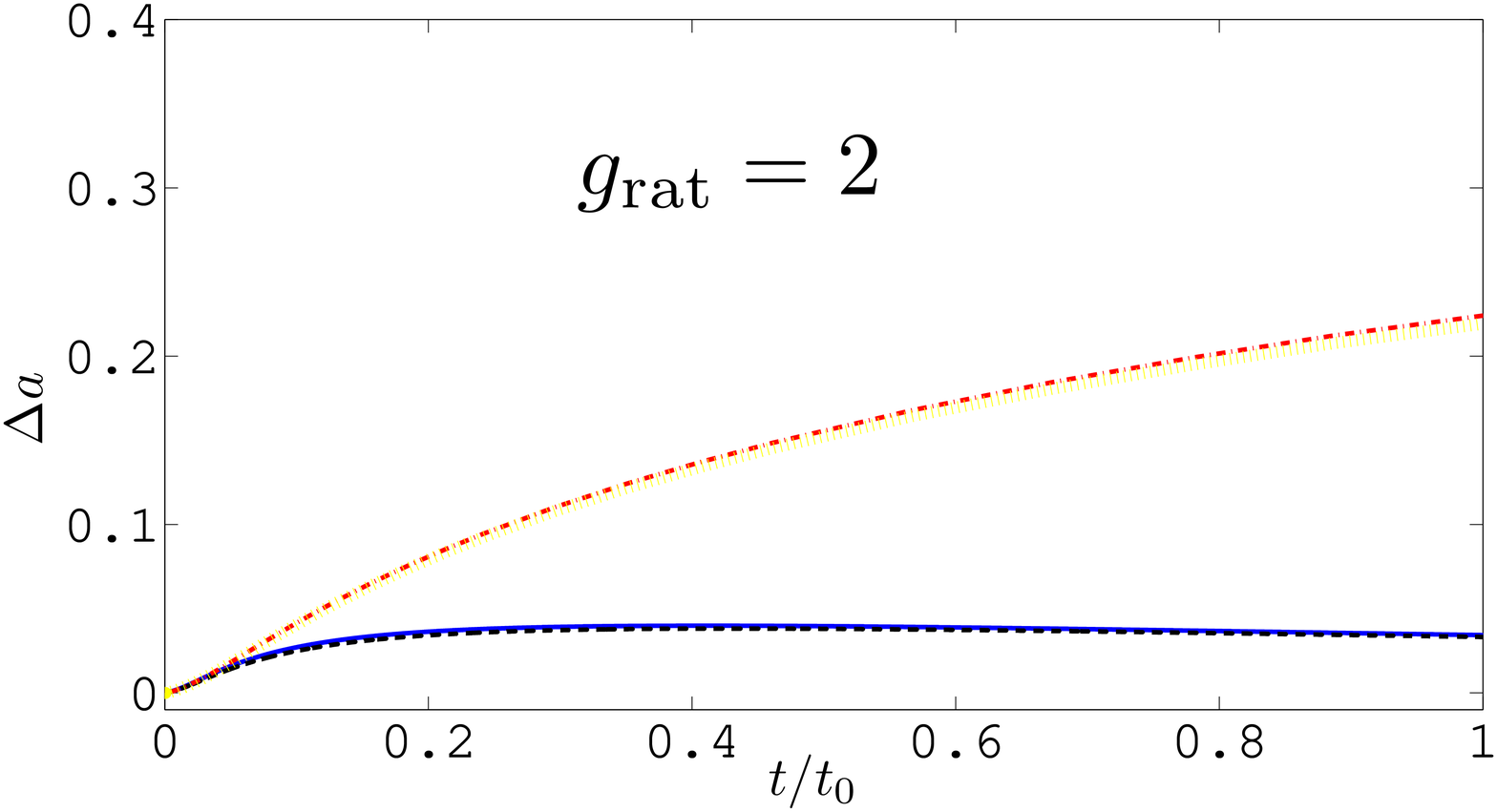}
\includegraphics[scale=0.2]{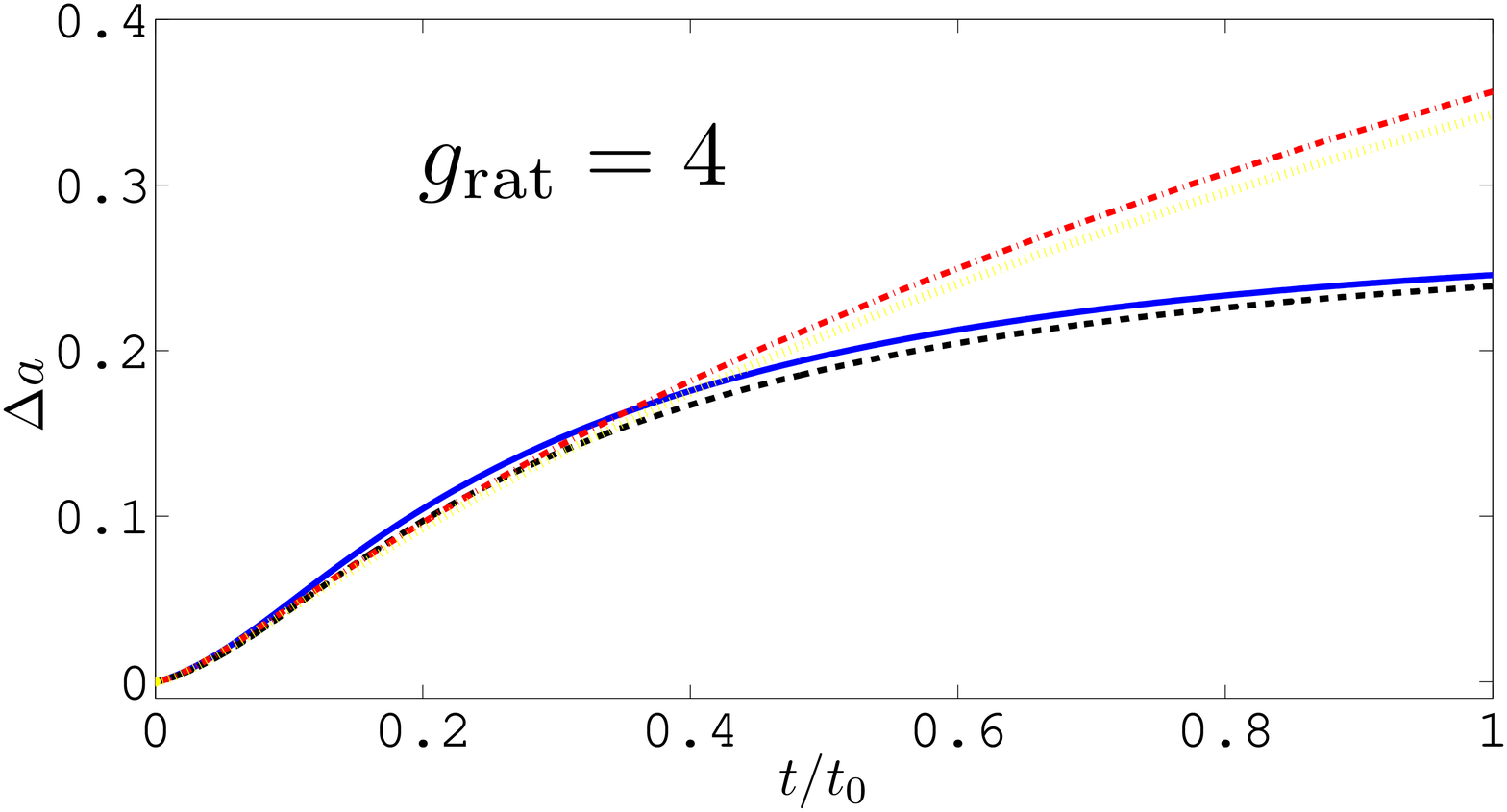}
\caption{\label{fig:delasphere}The relative difference between the average scale factor, $a_D$, and the scale factor of the background model, $(t/t_0)^{2/3}$, is plotted against time for the model presented in section \ref{sec:sphregions}. This model consists of a grid of spherically symmetric regions, where $\Phi_{,ii}=\Phi_{,11}$. $\Phi_{,11}$ is generated from a nearly scale-invariant, Gaussian, power-law, primordial spectrum. All three plots use the unsquared, metric in eq.\eqref{metric-1-phi}. The metric has either been truncated after including one gradient term (i.e. $\Phi_{,ij}$) or two. The upper panel has no fix on expanding regions. The lower two panels do, with thresholds of $g_{\rm rat}=2$ and $4$ respectively (see section \ref{sec:fixvoid} for details). The $a_D$ curves use the actual average scale factor defined in eq.\eqref{eq:defaD}. The $a(Q)$ curves use an average scale factor reconstructed from $Q$ using eq.\eqref{eq:Qfroma} (see section \ref{sec:aQdefs} for details). The grid spacing used in this plot is $R=1 \, {\rm Mpc}/h$ and the number of grid points is $N_x^3=30^3$ (see section \ref{sec:byeUV}). $t_0$ is the time today.}
\end{figure}
\paragraph{Figure \ref{fig:delasphere}:} We have plotted
\beq
\Delta a = \frac{a_D-\left(\frac{t}{t_0}\right)^{2/3}}{\left(\frac{t}{t_0}\right)^{2/3}}
\eeq
for this model in figure \ref{fig:delasphere}. The grid spacing used for these curves is $R=1 \, {\rm Mpc}/h$ and the number of grid points is $N_x^3=30^3$.\footnote{This grid spacing was used because, as can be seen in figure \ref{fig:Omeg_X}, it is approximately this grid spacing that gives the maximum \emph{rate} of backreaction in the late universe.}  We have used both the definitions of $a_D$ presented in section \ref{sec:aQdefs} and have shown results when the metric is truncated to include one gradient term or two. Finally, we have shown the results when we vary the threshold, $g_{\rm rat}$, described in section \ref{sec:fixvoid}. As is expected from the earlier results in section \ref{sec:sphere}, the order of truncation does not have a significant effect on the results. For the case where there is no fix at all on the expanding regions (upper panel) all four methods to calculate $a_D$ give similar results. While this result is unphysical because the expanding regions will not really expand indefinitely without interaction, this result reassures us that at the very least we have the mathematics of our model under control.

The curves that result when we do fix the expanding regions to the background are also what we expect. After sufficient time has passed, the curves that use the actual average scale factor, $a_D$, stop increasing. This time corresponds to the point when the bulk of points in the grid have either collapsed or been fixed. At this point any backreaction at this scale ends and $\Delta a$ becomes constant. When the threshold is increased, this turnover occurs later. Also, due the non-dusty method of applying this fix, the curves that are obtained from $Q$ do not see the effective pressure and continue increasing. Note that even once $Q\rightarrow 0$, $a(Q)$ continues slowly increasing. This is because a continuous pressure is required to keep the expanding regions from expanding faster than the original background model.

We do not know what the correct $g_{\rm rat}$ value is for our universe. In an extension of our model that includes $\Lambda$, the $a_D$ curves in figure \ref{fig:delasphere} would serve as a prediction for the magnitude of backreaction expected in the universe as a function of $g_{\rm rat}$.

\begin{figure}[t]
\includegraphics[scale=0.2]{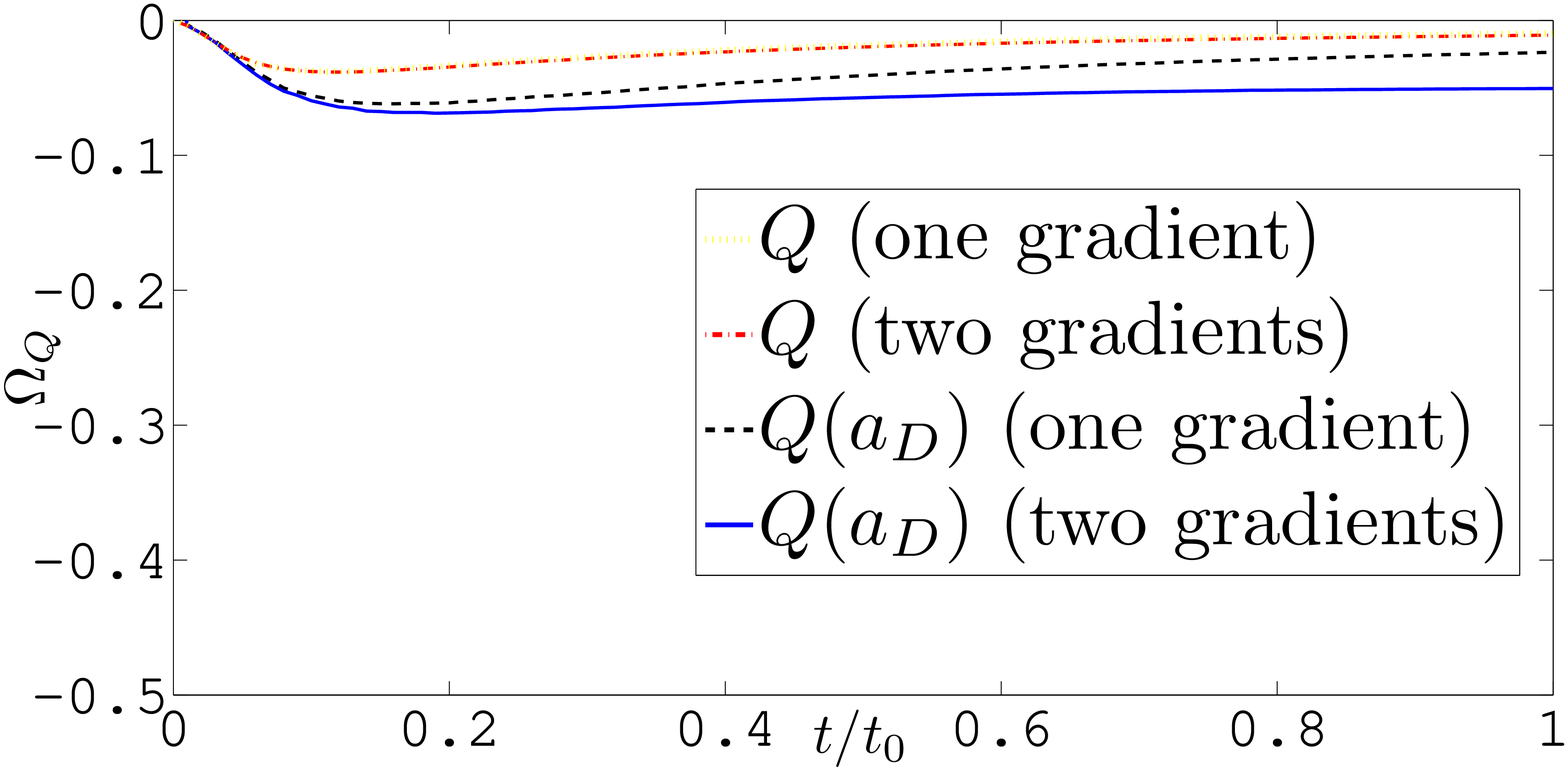}
\includegraphics[scale=0.2]{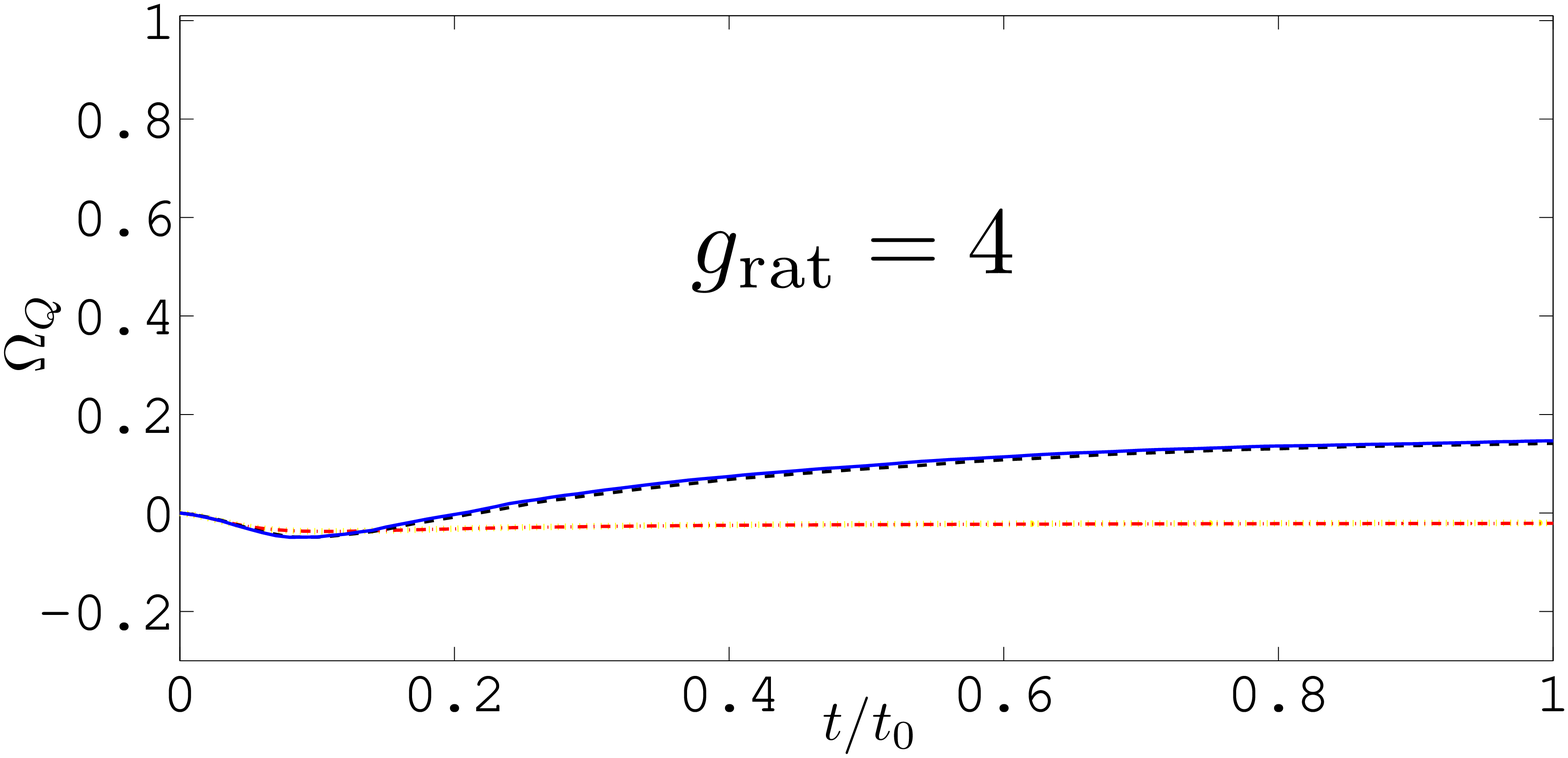}
\caption{\label{fig:Qsphere}$\Omega_Q$ is plotted against time for the model presented in section \ref{sec:sphregions}. The grid used is identical to the one used in figure \ref{fig:delasphere}. The left panel has no fix on expanding regions. The right panel does, with a threshold of $g_{\rm rat}=4$. The $Q$ curves use the actual value for $Q$ defined in eq.\eqref{eq:Qdef}. The $Q(a_D)$ curves use a $Q$ reconstructed from $a_D$ using eq.\eqref{eq:Qfroma} (see section \ref{sec:aQdefs} for details). $t_0$ is the time today.}
\end{figure}
\paragraph{Figure \ref{fig:Qsphere}:} We have plotted
\beq
\Omega_Q=\frac{-Q}{6 \left(\frac{\dot{a}_D}{a_D}\right)^2}
\eeq
for $Q$ and $Q(a_D)$ in figure \ref{fig:Qsphere} (see \cite{2008GReGr..40..467B} for a definition of $\Omega_Q$). The model is identical to that used in figure \ref{fig:delasphere}. The left panel has no fix on the expanding regions and the right panel uses $g_{\rm rat}=4$. We see the same behaviour in this figure as we did in figure \ref{fig:delasphere}. When there is no fix, all four methods for calculating $Q$ roughly coincide. When we do fix the expanding regions to the background, $Q$ is roughly unaffected. However $Q(a_D)$ changes significantly due to the effective pressure this fixing produces. The magnitude of ``pressure'' being applied could in fact be calculated from the asymptotic value of $\Omega_{Q(a_D)}$. Note that in all examples the maximum value of $\Omega_Q$ is $\gtrsim0.1$. Therefore, at least in this model, it appears that backreaction is large enough to have a measurable impact on dark energy constraints. Note here that even after $a_D$ has formed a plateau, $Q(a_D)$ is negative, this is because of the pressure required to keep the expanding regions expanding at the same rate as the background.

\subsubsection{Backreaction with the squared metric}\label{sec:squared}

Our initial motivation in using the gradient expansion was to construct a numerical model universe where the correlations between separate regions were taken into account. We deviated quite significantly from this motivation in the previous section by removing a lot of information from the initial metric. This was necessary because if we used all of the components in $\Phi_{,ij}$ the determinant of the metric $\gamma$ was unphysically skewed towards collapsing regions. We explained in that section that this occurred because one of the terms in the expression for $\gamma$ was unphysicallly skewed towards negative values as a result of the truncation of the metric.

In \cite{Croudace:1993yt} the authors performed a trick to avoid the determinant of this truncated metric crossing zero. We explain this trick and write the metric that it produces in appendix \ref{app:sqmet}. We will call the derived metric the ``squared metric'' because its derivation involves taking the square of the metric. The resulting metric contains at each order of the gradient expansion some extra higher order terms to ensure the positivity of the determinant. In \cite{Croudace:1993yt} it was found that this squared metric actually improved the match between the gradient expansion and the exact solution they were studying (the Szekeres solution). Unfortunately, as we show in appendix \ref{app:sqmet}, the squared metric matches our test case, the expanding and contracting spherical regions, worse.

Nevertheless, even at the first order of the expansion, the squared metric includes terms that involve the square of the gradient. Therefore the squared metric behaves much less pathologically. This is because, as well as the terms that are skewed towards negative values in eq.\eqref{eq:detform}, there will now include many other terms that are skewed positive. Whether a region collapses or gets fixed as an expanding region will depends on which terms are larger in that region. The non-pathological behavour of the squared metric allows us to use all of the components of $\Phi_{,ij}$ in the metric, thereby keeping all the information concerning the correlations between grid points. This also makes each individual grid point more realistic as it will not be isotropic and will expand and contract at different rates along different axes, which is exactly what happens in the real universe.

\paragraph{Fixing the squared metric:} There is a much more diverse range of behaviour with the squared metric. Some regions first expand, then contract, then expand again a number of times. For the one-gradient truncated metric, the asymptotic behaviour is however the same. That is, some regions end up expanding indefinitely and others collapse to a point.\footnote{Though note that if left alone, these ``collapsed'' regions would turn around and start to expand again which is an artifact of the squaring of the metric. This never happens in our model because we always freeze the evolution of any collapsed region.} We fix the expansion for this metric in exactly the same manner as before. However, for the two-gradient truncated metric, the asymptotic behaviour is different. No single region in the grid ever reaches a point of asymptotic expansion. This would be great if it were not for the fact that instead, all regions eventually collapse. This is clearly just as unphysical as every expanding region in the universe expanding indefinitely.

We still wish to fix the expansion of any over-expanded regions to the background. However we choose to take advantage of the extra complexity afforded to us by this metric. Given that every initially expanding region will eventually collapse, we fix the expanding regions and freeze the collapsing regions in a two step process. Firstly, exactly as before, we set a density threshold for collapsed regions, $\rho_{\rm rat}$, and a volume threshold for expanding regions, $g_{\rm rat}$. Once any gridpoint satisfies either the condition in eq.\eqref{eq:collapse} or \eqref{eq:expand} we peg that region as either a collapsing region or an expanding region respectively. However, we do not yet alter its evolution in any way. Whenever the density at a collapsing grid point stops expanding more slowly than the background and begins to expand faster than the background, we freeze it. Equivalently, whenever an expanding grid point turns over and begins to expand more slowly than the background, we fix it to the background. That is for a region labelled as collapsing we freeze the metric when
\beq
\frac{\gamma(t)}{t^4}>\frac{\gamma(t-\Delta t)}{(t-\Delta t)^4}
\eeq
is first satisfied. And for a region labelled as expanding we fix the metric to the background when
\beq
\frac{\gamma(t)}{t^4}<\frac{\gamma(t-\Delta t)}{(t-\Delta t)^4}
\eeq
is first satisfied. Note that these expressions should be understood to be evaluated in the limit that $\Delta t\rightarrow0$. Note again that once these points are reached, the method we use to freeze and fix the metric is identical to that described in section \ref{sec:fixvoid} and appendix \ref{app:smooth}.

\begin{figure}
\includegraphics[scale=0.4]{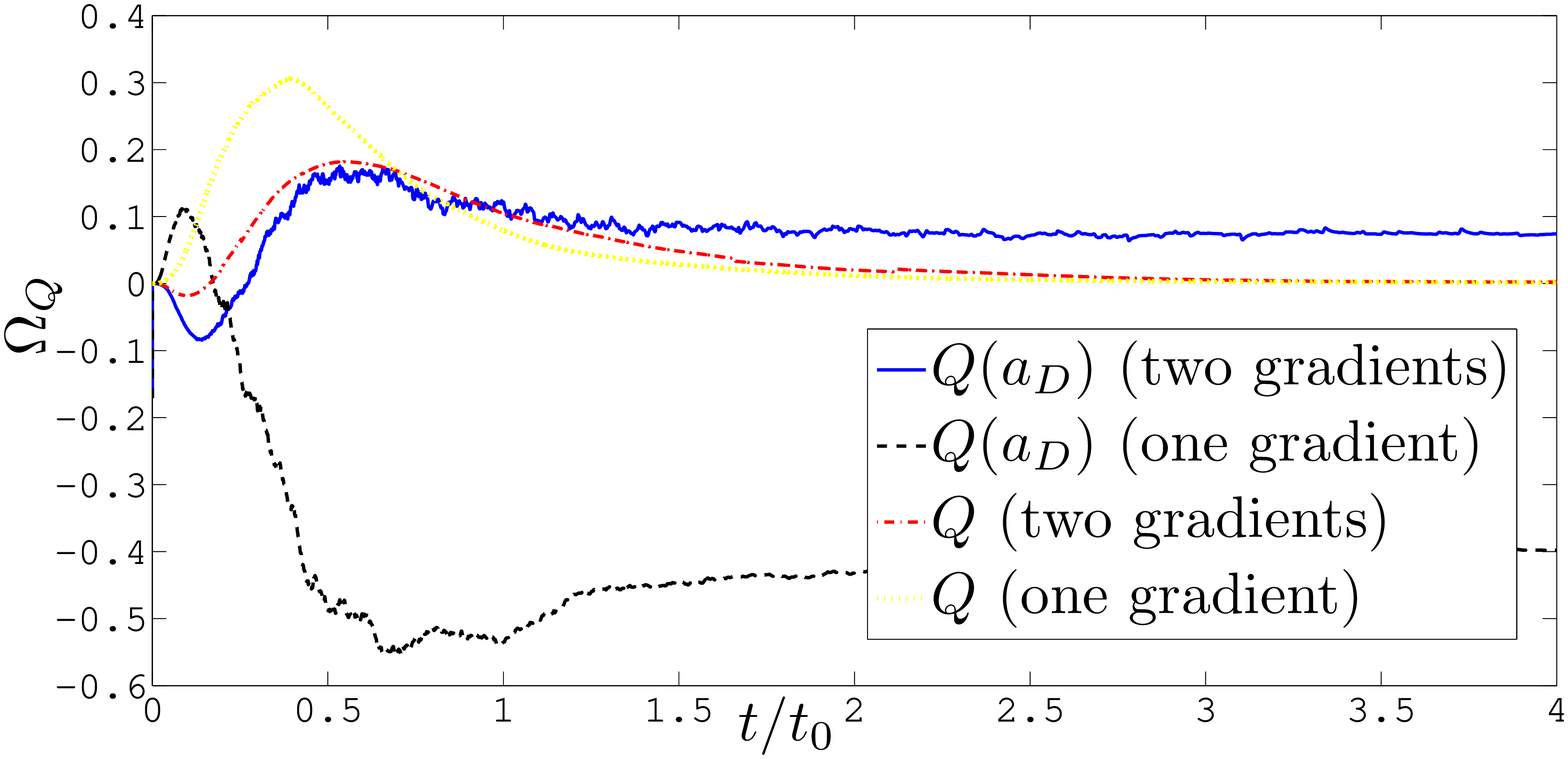}
\caption{\label{fig:Qsquare}$\Omega_Q$ is plotted against time for the model presented in section \ref{sec:squared}. This model uses all the components of $\Phi_{,ij}$, which is generated from a nearly scale-invariant, Gaussian, power-law, primordial spectrum. The metric used is the ``squared metric'' in eqs.\eqref{metric-3-phi} and \eqref{metric-5-phi}. The metric has either been truncated after including one gradient term (i.e. $\Phi_{,ij}$) or two. The curves all use thresholds of $g_{\rm rat}=3$ and $\rho_{\rm rat}=3$ for the fixing and freezing of expanding and collapsing regions respectively (see sections \ref{sec:fixvoid} and \ref{sec:squared} for details). The $Q$ curves use the actual value for $Q$ defined in eq.\eqref{eq:Qdef}. The $Q(a_D)$ curves use a $Q$ reconstructed from $a_D$ using eq.\eqref{eq:Qfroma} (see section \ref{sec:aQdefs} for details). The grid spacing used in this plot is $R=1 \, {\rm Mpc}/h$ and the number of grid points is $N_x^3=30^3$ (see section \ref{sec:byeUV}). $t_0$ is the time today.}
\end{figure}

\paragraph{Figure \ref{fig:Qsquare}:} We have plotted $\Omega_Q$ against time for the model universe generated using the squared metric. For these curves we have used $\rho_{\rm rat}=3$ and $g_{\rm rat}=3$. $R=1 {\rm Mpc}/h$ as usual. What is most interesting about these curves is that, for the truncated metric with two gradient terms, $Q(a_D)$ and $Q$ follow each other quite closely. This is a very nice result. The same is not also true for the truncated metric with one gradient term; however this is not surprising. Remember, the one gradient metric is fixed in exactly the same way as it was for the curves in figure \ref{fig:Qsphere}. As a result the same type of effective pressure will be present causing these two curves to deviate. The fact that the two-gradient curves match indicates that in that model of the universe, the universe remains dust dominated, even when we fix the metric and there is substantial backreaction.

The blue (solid) and red (dashed) lines therefore correspond to our best estimate of the backreaction in a true, statistically homogeneous, CDM universe that began flat. In this model, the backreaction generated at a given scale is first positive and then negative. These periods correspond to an initial over-expansion of under-dense regions, followed by an inhibition of these regions' growth which occurs initially as a natural result of the metric and subsequently due to our fixing the expansion to the background. The asymptotic value for both $Q$ and $Q(a_D)$ is approximately zero, indicating that eventually the backreaction generated at a given scale dies away.

The magnitude of $\Omega_Q$ can be as large as $\sim 0.2$. Using different grid spacings gives qualitatively identical results, except for a rescaling of the time axis. Altering $\rho_{\rm rat}$ and $g_{\rm rat}$ can have a significant effect on $|\Omega_Q|$; however it maximum value is always at least $\sim 0.1$.

\begin{figure}[t]
\includegraphics[scale=0.4]{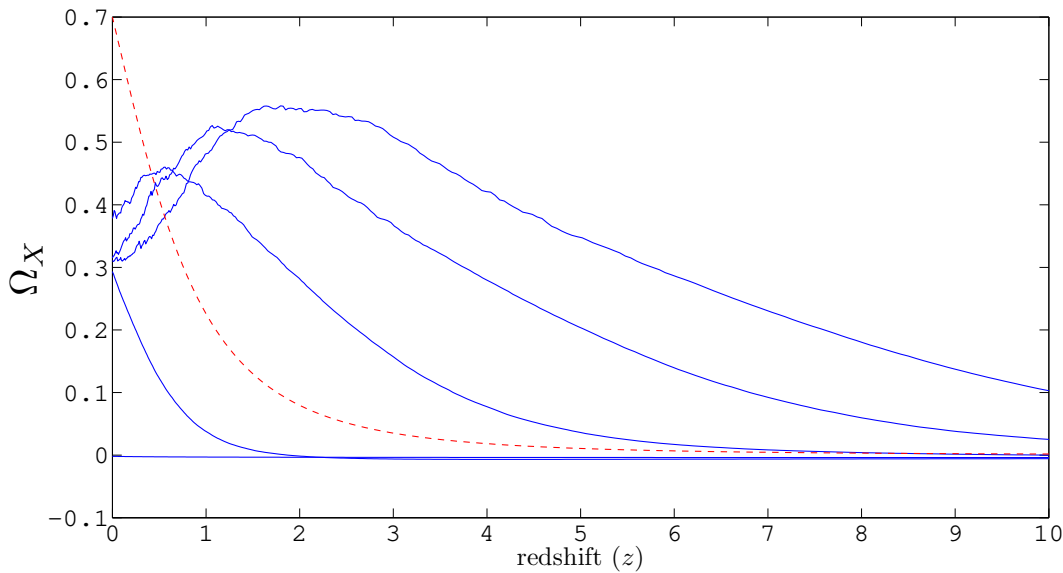}
\caption{\label{fig:Omeg_X}The cumulative backreaction $\Omega_X$ is plotted against redshift for the model presented in section \ref{sec:squared}. From right to left the solid blue curves correspond to grid spacings of $R=0.01\, {\rm Mpc}/h$, $0.1\, {\rm Mpc}/h$, $1 \,{\rm Mpc}/h$, $10\, {\rm Mpc}/h$ and $100\, {\rm Mpc}/h$ (in each case $N_x=20$). All of the curves use the squared metric, truncated after including two gradient terms. The curves all use thresholds of $g_{\rm rat}=3$ and $\rho_{\rm rat}=3$ for the fixing and freezing of expanding and collapsing regions respectively (see sections \ref{sec:fixvoid} and \ref{sec:squared} for details). The dotted red line is the expectation in a $\Lambda$CDM model with no backreaction and $\Omega_\Lambda=0.7$ today.}
\end{figure}

\paragraph{Figure \ref{fig:Omeg_X}:}In a universe with significant backreaction the averaged Friedman equation also contains a term from the averaged spatial curvature $\langle R\rangle$ - see e.g.\cite{2008GReGr..40..467B}.\footnote{Thanks to Thomas Buchert for pointing this out.} This term depends on the cumulative effects of backreaction; therefore it can be quite large and can contribute at least as much as $Q$ to the measured effects of backreaction today. In this section, we display the total, \emph{cumulative} backreaction.

The total contribution of backreaction to the average expansion rate can be encapsulated in one term, $\Omega_X$ defined through $\Omega_X=\Omega_Q+\Omega_R$. Note that $\Omega_R$ is defined in the same way as $\Omega_Q$ using the average expansion rate. Unfortunately, to calculate $\langle R\rangle$ at $t>t_i$ it is necessary to calculate second order spatial derivatives of the metric. This would involve calculating fourth order derivatives of the gravitational potential term that we use to generate the initial condition for the metric. We chose to circumvent this difficulty by directly comparing the density of matter to the expansion rate. This is possible because, in an initially flat CDM universe,
\begin{equation}
 \Omega_X(t)=1-\Omega_m(t)
\end{equation}
where $\Omega_m(t)$ is defined by the \emph{average} density and \emph{average} expansion rate at time $t$.

In figure \ref{fig:Omeg_X} we plot this $\Omega_X$ against ``average'' redshift, defined through $a_D=1/(1+z)$. We also show in this figure the effects of changing the grid spacing. Each curve in this plot uses the same values of $\rho_{\rm rat}$ and $g_{\rm rat}$ and is constructed using the squared metric truncated after including two gradient terms. It is clear that a smaller grid spacing induces backreaction at earlier times. One interesting feature of this plot is that while scales of approximately $1-10 \,{\rm Mpc}/h$ are showing relatively large backreaction today, when averaged over $100 \,{\rm Mpc}/h$ scales the universe is clearly still very homogeneous with small backreaction. From the arguments made in section \ref{HJ} we expect that up to $t/t_0=1$ ($z=0$) the region below the 10 $h^{-1}$ Mpc line is a realistic expectation for the magnitude of backreaction \emph{from such scales alone}. However, this plot does indicate that a statistically homogenous universe does not preclude the possibility of significant backreaction arising from structure growth at smaller scales. For reference, the dotted red line is the expected curve in a $\Lambda$CDM model with no backreaction and $\Omega_\Lambda=0.7$ today.

The fact that a range of scales that vary by more than an order of magnitude can all be backreacting at the same time indicates that our use of one fixed grid to assess the total backreaction in our model universe is lacking. If a truncated gradient expansion is to be used, then to properly quantify the total backreaction occurring in the universe at any given time it will not be sufficient to focus only on a small range of scales. How to tackle this problem is an open question, although we suggest some possibilities in section \ref{sec:extensions}.

\subsubsection{What our model universes look like}\label{sec:univsli}

\begin{figure}[t]
\includegraphics[scale=0.65]{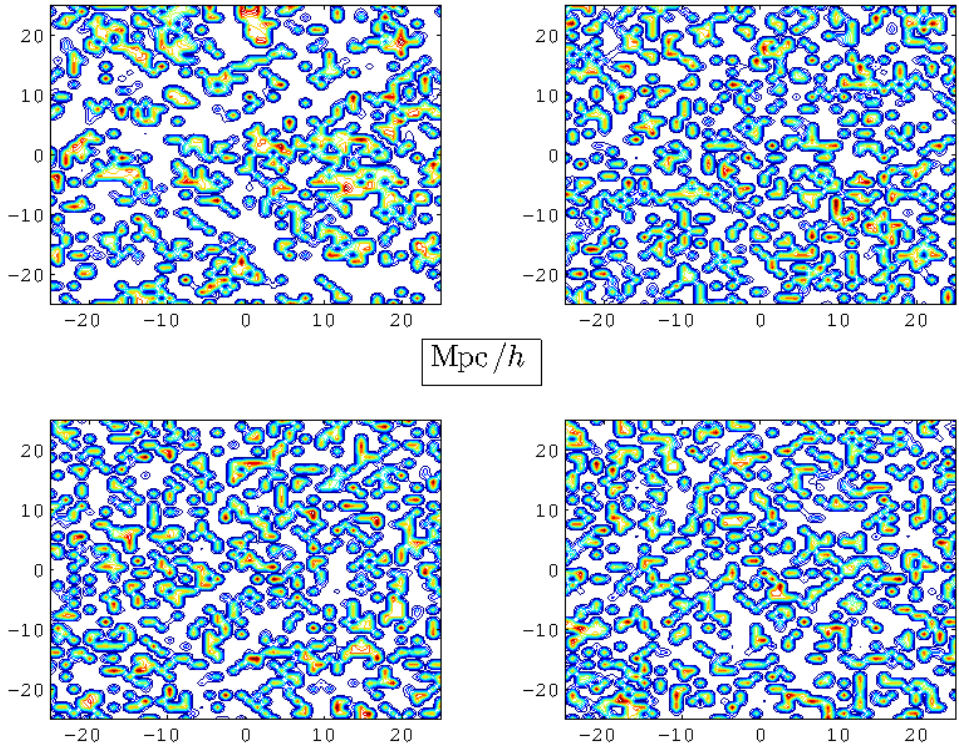}
\caption{\label{fig:volslice}This figure depicts the volume at each gridpoint in a two-dimensional slice of our $N_x^3$ grid at $t=t_0$. The upper left panel is an actual slice from our model, the other three panels are the result of redistributing each gridpoint to a new gridpoint using a completely random permutation. The axes show the \emph{synchronous comoving coordinates} of each gridpoint, which therefore \emph{do not} define physical distances at $t_0$. Instead, the volume of each gridpoint is depicted by colour. The white regions are made of collapsed gridpoints with effectively zero volume. The blue and red (dark) regions are all expanding regions. The more red a gridpoint is the bigger the volume. This figure was produced using the model described in section \ref{sec:squared}. The list of parameters used for this figure appears in section \ref{sec:univsli}.}
\end{figure}

All of the figures we have presented so far have quantified in some way the magnitude of backreaction in our models; however they do not show very well what the model universe actually \emph{looks} like. The initial conditions of the model include correlations between grid points, but the subsequent behaviour of each gridpoint is completely independent. The gradient expansion is an iterative solution to the full equations of general relativity, so a complete summation of this expansion should still describe the behaviour of each gridpoint exactly. However it is an important question to ask whether our truncated expansion retains information about the correlation between points, even after the terms in the expansion become large.

\begin{figure}[t]
\centering\includegraphics[scale=0.45]{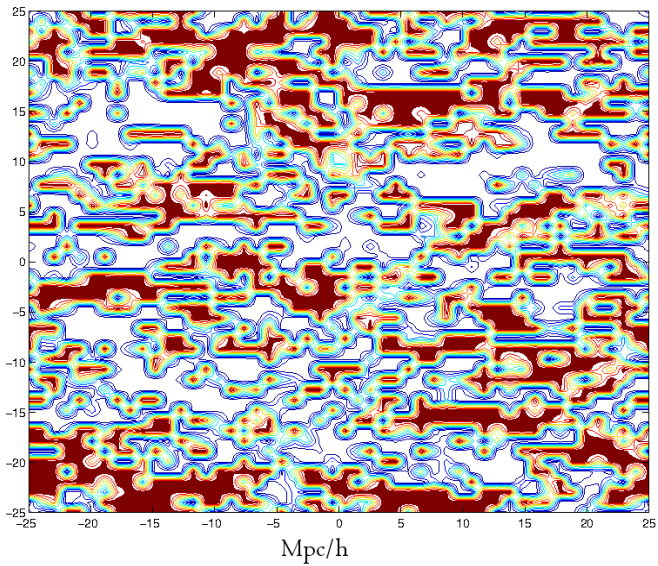}
\caption{\label{fig:volsph}This figure depicts the volume at each gridpoint using the same conventions as figure \ref{fig:volslice}. The only difference is that this figure was produced using the model described in section \ref{sec:sphregions}. The list of parameters used for this figure also appears in section \ref{sec:univsli}. See section \ref{sec:univsli} for a discussion of both figures.}
\end{figure}

In figures \ref{fig:volslice} and \ref{fig:volsph} we depict the volume of a two-dimensional slice of our grid for the models of sections \ref{sec:squared} and \ref{sec:sphregions} respectively. The gridpoints are defined by synchronous comoving coordinates therefore the axes in these figures and the width of each region do not depict the area of each region. Instead, we depict the volume of each gridpoint using colour. The white regions consist of all the collapsed gridpoints. The blue and red regions are the gridpoints that have been labelled as expanding. The more red a coloured region is the bigger its volume is. For each figure we use a gridspacing of $R=1\, {\rm Mpc}/h$ and $N_x=50$ gridpoints in each direction. We also use $\rho_{\rm rat}=g_{\rm rat}=3$ for each figure. The volume of each gridpoint is defined simply as the value of $\sqrt{\gamma}$ at that gridpoint. Finally, both figures are shown at $t=t_0$.

In figure \ref{fig:volslice} we show a slice of the grid for the model in section \ref{sec:squared}. This is shown in the upper left panel of this figure. In the other three panels we have redistributed each grid point using a random permutation. It is reasonably clear from this figure that the structures in our model universe look different to structures in a completely random collection of over and under-dense regions. The collapsed regions clump together more, as do the over-expanded regions. It is even arguable that a few filaments of collapsed regions can be made out in the slice from the real model.

In figure \ref{fig:volsph} we show a slice of the grid for the model in section \ref{sec:sphregions}. There is one striking feature of this figure that is immediately obvious, even to the human eye. There is a very clear preferred axis in this model. Almost all of the collapsed regions and expanding regions align along this axis. This is actually not surprising and in fact shows very clearly that correlations in our model do survive to late times. In constructing this model we set $\Phi_{,ii}=\Phi_{,11}$. Therefore we lost all information about correlations in the $x_2$ and $x_3$ directions, but not in the $x_1$ direction. It is the $x_1$ axis that the structures line up along. One other feature of this figure is the greater uniformity in the volume of the expanding regions. This is also not surprising given how the expanding regions are fixed to the background in this model. In the model of section \ref{sec:squared} we first label a gridpoint as expanding and then fix it only when its expansion slows to be the same as the background expansion. Different gridpoints will turn around when they have different volumes. In the model of section \ref{sec:sphregions} (used in this figure) the expanding regions are fixed as soon as they cross a certain threshold. The result of this and the method we use to do the fixing is that every region asymptotes to the same volume. The small differences that are seen in figure \ref{fig:volsph} result from some regions not having yet reached the asymptotic metric.

The fact that non-random structures are clearly visible in our models suggests that the of use a truncated gradient expansion to describe the universe is not completely without merit. At the very least, the existence of these correlated structures is one step better than earlier models that are made up of completely independent regions. Our models lack the sophistication of the best Newtonian N-body simulations. However backreaction is a purely relativistic effect. Therefore our approach could at present be considered as the most sophisticated one. We do not require any special symmetries of the inhomogeneities but can nevertheless explicitly calculate the backreaction.

\section{Discussion}

\subsection{Potential improvements to the model}\label{sec:extensions}

While our model is interesting and internally consistent and perhaps even an improvement in sophistication compared to earlier models, what we want is to know is the magnitude and nature of the backreaction in the real universe. The most obvious next step in improving our model is to add a cosmological constant term, $\Lambda$. With that achieved, it would be possible to do a full Markov Chain Monte Carlo over the parameters of the model to see which set of parameters fits the observed data best. If our model, with backreaction, fitted the data better than the concordance model, without backreaction, then this would indicate that the assumptions we made to solve the various complications were well-motivated.

However, another complication that is less easily solved would need to be addressed before comparisons to the real universe can be meaningfully made. In this paper we have calculated the backreaction that arises as structures form over a small range of scales. We can vary where this range lies, but we cannot make it arbitrarily wide. To compare the model to observation we need a reliable calculation of the \emph{total} backreaction that arises over \emph{all} scales. In principle this problem can be solved by considering a larger box; however, in practical terms, this can never be fully achieved.

Given the hierarchical nature of structure formation, an ideal solution to this problem is a time-dependent grid spacing and box size. The time-dependence could be global, with the grid spacing $R$ set as a function of time $R(t)$. $R(t)$ would be chosen based on which scales start forming structures at which times and at which times structure growth freezes out. Alternatively, the time-dependence could be set locally. There could be a progressive ``zooming out'' as individual gridpoints expand into neighbouring gridpoints. Whenever two gridpoints collide it would be possible to stop considering them separately and to recalculate the metric from that time onwards as if they were one gridpoint. Precisely what metric to use for this combined gridpoint would need to be determined. Either a global or a local time-dependence to the grid spacing would face this difficulty. That is, the question of precisely what initial conditions, or background model, to use for the zoomed out metric. Should $t_0$ (which was set as a function of $h$) become a time-dependent quantity? If so, should the $h$ used in the transfer function also be dependent on the grid spacing and thus time?

If the zooming out problem could be solved it would remove any problems that relate to how to fix the over expansion of gridpoints. However, if the zooming out problem is not solved, there are some improvements that can be made to our fixing methods. Firstly, it should be possible to pick $g_{\rm rat}$ from observations, rather than to treat it as a free parameter, as we have here. Secondly, we chose to fix our expanding regions to the initial background model. This is not ideal. When any region reaches the point where we wish to fix its expansion, the \emph{average} expansion rate, and more importantly, the average volume and density, are not the same as those of the initial background, extrapolated to this time. If the universe is to reach a new asymptotic FRW state it should be related to the new average density, not the one extrapolated from the initial background model. However, doing this is not easy. The deviation of the average density from the initial background model changes with time. Therefore, the correct background model to fix the expanding regions to also changes with time. When the total backreaction is $<10\%$ these issues will be less important than the many other features of the model. However, when comparing the model to precise measurements, perhaps $10\%$ of $10\%$ is no longer an ignorably small quantity.

\subsection{Conclusions}

Let us close by summarizing our methodology and the conclusions we can draw from it. We have approximated the evolving inhomogeneous synchronous gauge metric of a CDM $\Omega=1$ universe with realistic initial conditions using a gradient expansion, and studied the backreaction on the average dynamics. This gradient expansion is an approximation to the true metric of such a universe which goes beyond the description of standard perturbation theory. Instead of being limited by the smallness of the inhomogeneities, it has a temporal range of validity set by the initial local 3-curvature. For timescales $t>t_{\rm con}$ the gradient expansion does not converge anymore. This does not pose a real problem for studying regions with positive initial curvature which collapse at around that timescale and eventually virialize. However, it does present limitations in following the late time evolution of under-dense regions with negative initial curvature. Nevertheless, if the late time evolution of such regions is fixed in some prescribed manner, this approach offers a way to construct a realistic \emph{model} even for the late-time metric of such a universe.

Our methodology improves over previous approaches in that it is fully relativistic, goes beyond perturbation theory and does not treat the universe as a collection of unconnected under-dense and over-dense structures with high symmetry. Of course it does require modeling of the late time behaviour of the under-dense regions which is not naturally captured by the gradient expansion but such modeling can be simply parameterized and possibly informed by observation. We have attempted to achieve this in a simple manner in this paper but it can still be made more realistic in a number of ways.

Given the approximations mentioned, we computed the backreaction of inhomogeneities in this model metric. Our qualitative results indicate that backreaction may constitute more than a percent effect and is thus highly relevant in future considerations of precision cosmology. Of course more realism is needed before definite statements can be made regarding the backreaction in our universe.

There is another aspect in which such a model would aid in understanding the effects of inhomogeneities for cosmology. We have focused on the average dynamics but what we actually see is light propagating through the inhomogeneous universe and not the average scale factor. One could question whether the latter is at all relevant for our observations. To really assess the impact of inhomogeneities on observations one should trace light rays through the inhomogeneous spacetime and determine what observers would see \cite{arXiv:1002.1232,arXiv:1109.2484}. This is easy to do in a box described by our metric. In particular we can compute what happens to the redshift of photons or the travel time though over-dense and under-dense regions. The great advantage of our approach is that this is easily calculable within our model. Therefore, even irrespective of whether our model describes the universe entirely accurately, we can answer the question of whether a large backreaction in the synchronous gauge corresponds to a large shift in the time it takes a photon to traverse the universe and/or the redshift it experiences. We will return to this issue and further improvements to our model in future work.

\section{Acknowledgements}
We wish to thank Thomas Buchert for his valuable input. KE is supported by the Academy of Finland grant 218322. SH is supported by the Academy of Finland grant 131454. GR is supported by the Gottfried Wilhelm Leibniz programme of the Deutsche Forschungsgemeinschaft.

\appendix

\section{The squared metric}\label{app:sqmet}

The metric (\ref{metric-1-phi}) is the result of a gradient expansion approximation to the Einstein equations for a dust dominated universe. It has however the undesirable feature that this expression eventually becomes negative in regions of positive curvature. To overcome this, the authors of \cite{Croudace:1993yt} advocated the following fix: they took the square root of the original metric to the desired gradient order and then squared it again. More concretely, in our case this procedure yields for the 1st and 2nd order metric

\beq \label{metric-3-phi}
\gamma_{ij}=\left(\frac{t}{ t_0}\right)^\frac{4}{3}\left[\delta_{il}
+\frac{3}{2}t^{2/3}{t_0}^{4/3}\Phi_{,il}\right]\left[\delta^l_{j}
+\frac{3}{2}t^{2/3}{t_0}^{4/3}\Phi^{,l}{}_{,j}\right]
\eeq

\beq\label{metric-5-phi}
\gamma_{ij}=\left(\frac{t}{ t_0}\right)^\frac{4}{3}\left[\delta_{il}
+\frac{3}{2}t^{2/3}t_0^{4/3}\Phi_{,il}+\frac{1}{2}t^{4/3}t_0^{8/3}\hat{B}_{il}\right]\left[
\delta^l_{j}
+\frac{3}{2}t^{2/3}t_0^{4/3}\Phi^{,l}{}_{,j}+\frac{1}{2}t^{4/3}t_0^{8/3}\hat{B}^{l}{}_{j}
\right]
\eeq
where $B_{ij}$ is now
\beq
\hat{B}_{ij}=\frac{27}{28}\left[4\left(\Phi_{,il}\Phi^{,l}{}_{,j}-\Phi_{,ij}\Phi^{,l}{
}_{,l}\right)
+\delta_{ij}\left((\Phi^{,l}{}_{,l})^2-\Phi_{,lm}\Phi^{,lm}\right)\right]\,.
\eeq
Note that now there are higher order terms which ensure that the metric is always positive.

\begin{figure}[t]\label{sphericalsquared}
\includegraphics[scale=1]{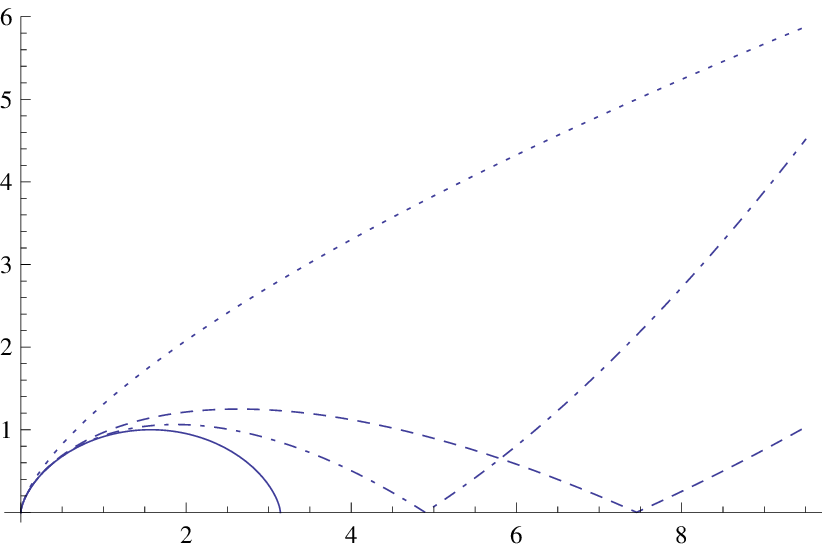}
\includegraphics[scale=1]{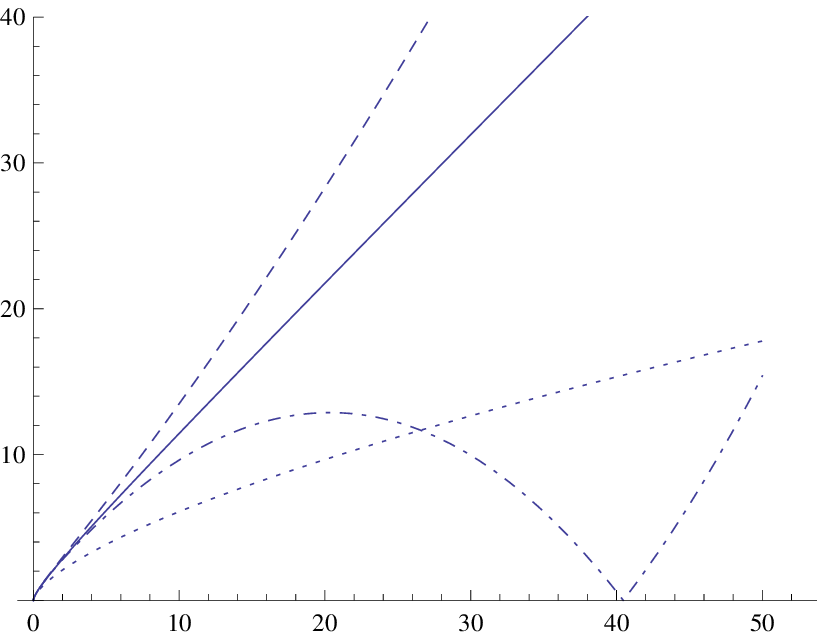}
\caption{The "squared metric" for the same configuration as in figure \ref{spherical}. The solid line is the exact solution, the dotted line is the zero gradient approximation for the metric (flat FRW or "separate universe approximation"), the dashed line is the 1st order metric (2 spatial derivatives) squared and the dash-dotted line is the 2nd order metric (4 spatial derivatives) squared. The difference in the behavior compared to the straightforward gradient expansion for large times is obvious.}
\end{figure}

It was argued in \cite{Croudace:1993yt} that such a procedure improves convergence and in fact can reproduce exactly the Szekeres solution studied there. When applied to our spherical test case it actually does worse than the straightforward gradient expansion as can be seen from figure \ref{sphericalsquared}. However, its use for studying backreaction has merits, as argued in the main text. Furthermore, we see that the (less important for backreaction) collapsing regions are still modeled relatively well while the behaviour of the expanding regions at two gradients gets regulated since they eventually slow down. In this configuration our recipe for fixing the squared metric would result in the collapsing region shrinking down to zero volume (and hence be irrelevant for backreaction) while the expanding region would be made to follow the background evolution at $t\sim 14$. This corresponds to a volume ratio between the expanding region and the background $V_{\rm exp}/V_{\rm bg}\sim 3.5$.

\section{Fixing the metric continuously}\label{app:smooth}

To avoid sharp features appearing in $Q(a_D)$ (and to a lesser extent $Q$) it is advantageous to smoothly impose the fix of the expanding regions to the background. For the model that consists of spherically symmetric regions, introduced in section \ref{sec:sphregions}, the asymptotic state that is required is written in eq.\eqref{eq:voidfix}. We choose to to impose the following smooth transition

\begin{eqnarray}\label{eq:fullvoidfix}
\gamma_{ii}(t)&=&\gamma^{1/3} \left(\frac{t}{t^*}\right)^{4/3}-\frac{1}{\alpha}e^{-\alpha(t-t^*)}\left(\dot{\gamma}_{ii}(t^*)-\frac{4\gamma^{1/3}}{3t^*}\right)+
\frac{1}{\alpha}\left(\dot{\gamma}_{ii}(t^*)-\frac{4\gamma^{1/3}}{3t^*}\right)
\nonumber \\
\dot{\gamma}_{ii}(t)&=&\frac{4\gamma^{1/3}}{3t^*}\left(\frac{t}{t^*}\right)^{1/3}+e^{-\alpha(t-t^*)}\left(\dot{\gamma}_{ii}(t^*)-\frac{4\gamma^{1/3}}{3t^*}\right)\,\, ,
\end{eqnarray}
imposed for all $t>t^*$, which asymptotically approaches eq.\eqref{eq:voidfix}. Note, $\gamma=\gamma(t^*,{\bf x}^*)$. Remember that for the model of spherically symmetric regions, $\gamma_{11}=\gamma_{22}=\gamma_{33}$ and $\gamma_{ij}=0$ for all $i\neq j$. Note that at $t=t^*$, $\gamma_{ii}=\gamma^{1/3}$ and $\dot{\gamma}_{ii}=\dot{\gamma}_{ii}(t^*)$ as is required for the transition to be smooth. Although $\ddot{\gamma}_{ii}$ (and therefore $\ddot{a_D}$) will not be continuous during this transition, it is well-defined. By increasing $\alpha$ the transition can be made sharper. This will result in sharper features in $Q(a_D)$. By making $\alpha$ smaller the transition can be made to be more gradual. However this has the effect of increasing the final constant term in the expression for $\gamma_{ii}$ in eq.\eqref{eq:fullvoidfix} making the grid point take much longer to reach the asymptotic state.

For the model introduced in section \ref{sec:squared}, the issue of smoothness in the transition is not as important. That model uses the squared metric and all of the components in $\Phi_{,ij}$. For that model, we actually impose the fix to the background at precisely the moment when $\dot{\gamma}$ at some grid point is the same as the background. That is, the moment when the gridpoint stops expanding faster than the background. We still apply the fix described in \eqref{eq:fullvoidfix}, with the additional conditions that $\gamma_{ij}=\dot{\gamma}_{ij}=0$ for all $i\neq j$ (and all $t>t^*$). This instantly forces the gridpoint into a spherically symmetric state, which is obviously not a smooth or even continuous transition; however $\gamma$ and $\dot{\gamma}$ do remain continuous and therefore so do $a_D$ and $\dot{a}_D$.

For the two lower panels of figure \ref{fig:delasphere} we used $\alpha=1$. For figure \ref{fig:Qsquare} we usd $\alpha=50$. For figure \ref{fig:Omeg_X} we used $\alpha=20$. For figure \ref{fig:volslice} we used $\alpha=5$. And for figure \ref{fig:volsph} we used $\alpha=1$. The much larger values of $\alpha$ used for the model of section \ref{sec:squared} is a result of the transition already being smooth. That is, using eq.\eqref{eq:fullvoidfix} instead of eq.\eqref{eq:voidfix} does not improve the situation. Setting $\alpha\gg1$ effectively makes the transition instantaneous. However setting $\alpha\gg1$ for the model of section \ref{sec:sphregions} makes $Q(a_D)$ much more noisy.

\end{document}